\title{Integration over Spin-Angular Variables in Atomic Physics}
\author{ Gediminas Gaigalas\\
{\em Institute of Theoretical Physics and Astronomy,}\\
{\em A. Go\v stauto 12, Vilnius
2600, LITHUANIA}}
\documentclass[11pt]{article}
\usepackage{graphicx}
\begin{document}
\maketitle

%\pagenumbering{arabic}

\vspace{.5in} {\bf PACS: 0270, 3110, 3115}

%\clearpage

\begin{abstract}

A review of methods for finding general expressions for
matrix elements (non-diagonal with respect to configurations included)
of any one- and two-particle operator for an arbitrary
number of shells in an atomic configuration is given.
These methods are compared in various aspects, and the advantages or
shortcomings of each particular method are discussed.
Efficient method to find the abovementioned quantities in $LS$ coupling
is presented, based on the use of symmetry properties of operators
and matrix elements in three spaces (orbital, spin and quasispin),
second quantization in coupled tensorial form, graphical technique and
Wick's theorem. This allows to efficiently account for
correlation effects practically for any atom and ion of periodical table.

\end{abstract}

\clearpage

\section{Introduction}

Modern atomic spectroscopy studies the structure and properties of
practically any atom of the periodic table as well as of ions of any
ionization degree. Particular attention is paid to their energy spectra. For
the investigations of many-electron atoms and ions, it is of great
importance to combine experimental and theoretical methods. Nowadays the
possibilities of theoretical spectroscopy are much enlarged thanks to the
wide use of powerful computers. Theoretical methods utilized must be
fairly universal and must ensure reasonably accurate values of physical
quantities studied.

Many-electron atom usually is considered as many-body problem and is
described by the wave function constructed from the wave functions of one
electron, moving in the central nuclear charge field and in the screening
field of the remaining electrons. Then the wave function of this electron
may be represented as a product of radial and spin-angular parts. The radial
part is usually found by solving various modifications of the Hartree-Fock
equations and can be represented in a numerical or analytical forms (Froese
Fischer~\cite{fa}) whereas the angular part is expressed in terms of spherical
functions. Then the wave function of the whole atom can be constructed in
some standard way (Cowan~\cite{Cowan}, Jucys and Savukynas~\cite{js},
Nikitin and Rudzikas~\cite{NR}) starting with these one-electron functions
and may be used
further on for the calculations of any matrix elements representing physical
quantities.

For obtaining the values of atomic quantities it is necessary to
solve so-called eigenvalue problem

\begin{equation}
\label{eq:Sch}H\Psi =E\Psi ,
\end{equation}
where $\Psi $ is the wave function of the system under investigation and $H$
is its Hamiltonian.
In order to obtain accurate values of atomic quantities it is necessary to
account for relativistic and correlation effects.
It turned out that for very large variety of atoms and their ionization degrees the
relativistic effects may be taken into account fairly accurately
as Breit-Pauli corrections
(Nikitin and Rudzikas~\cite{NR}, Rudzikas~\cite{r}).
It is convenient to present the Hamiltonian as consisting
of two parts in Breit-Pauli approach, namely,

\begin{equation}
\label{eq:ha}
H_{BP} = H_{NR} + H_{R},
\end{equation}
where $H_{NR}$ is the
non-relativistic Hamiltonian and $H_{R}$ stands for the relativistic
contibution.
Non-relativistic Hamiltonian is the sum of the kinetic energy of electrons
$T$, their potential energy $P$, and the electrostatic electron interaction
$Q$

\begin{equation}
\label{eq:hb}
H_{NR} = T + P + Q.
\end{equation}
The first two terms, $T$ and $P$ are the one-particle operators,
whereas the third term, $Q$, is two-particle operator.
The $H_{R}$ may also be subdivided into non-fine structure and fine
structure contributions

\begin{equation}
\label{eq:hc}
H_{R} = H_{Non-Fine} + H_{Fine}.
\end{equation}
The non-fine structure contributions

\begin{equation}
\label{eq:hd}
H_{Non-Fine} = H_{Mass} + H_{D1} + H_{D2} + H_{SSC} + H_{OO}
\end{equation}
shift non-relativistic energy levels without any splitting of them.
The mass-velocity term $H_{Mass}$ describes the variation of mass with velocity.
The one- and two-body Darwin (contact) terms $H_{D1}$ and  $H_{D2}$ are
the corrections of the
one-electron Dirac equation due to the retardation of the electromagnetic
field produced by an electron (contact interaction). The spin-spin-contact
term $H_{SSC}$ accounts
for the interaction of the spin magnetic moments of two electrons occupying
the same space. The orbit-orbit interaction $H_{OO}$ accounts for the
interaction of two orbital moments.

The fine-structure contributions

\begin{equation}
\label{eq:he}
H_{Fine} = H_{SO} + H_{SOO} +  H_{SS}
\end{equation}
split the non-relativistic energy levels (terms) into a series of
closely-spaced levels.
The most important of these is the nuclear spin-orbit interaction $H_{SO}$
(spin-own-orbit)
representing the interaction of the spin and angular magnetic moments of an
electron in the field of the nucleus. The spin-other-orbit $H_{SOO}$ and
spin-spin $H_{SS}$ contributions in a rough sense may be viewed as corrections
to the nuclear spin-orbit interaction due to the presence of the other
electrons in the system.

Unfortunately, practical calculations show that all realistic atomic
Hamiltonians do not lead straightforwardly to eigenvalue problem
(\ref {eq:Sch}).
Actually we have to calculate all non-zero matrix elements of the
Hamiltonian considered including those non-diagonal with respect to
electronic configurations, then to form energy matrix, to diagonalize it,
obtaining in this way the values of the energy levels as well as the
eigenfunctions (the wave functions in the intermediate coupling scheme). The
latter may be used then to calculate electronic transitions as well as the
other properties and processes. Such a necessity raises special requirements
for the theory.

The total matrix element of each term of the energy operator in the case of
complex electronic configuration will consist of matrix elements, describing
the interaction inside each shell of
equivalent electrons as well as between these shells. Going beyond the
single-configuration approximation we have to be able to take into account
in the same way non-diagonal, with respect to configurations, matrix
elements.

To find the
expressions for the matrix elements of all terms of the Hamiltonian
considered for complex electronic configurations, having several open
shells, is a task very far from the trivial one. A considerable part of the effort
must be devoted to coping with integrations over spin-angular variables,
occurring in the matrix elements of the operators under consideration.
This paper presents the general methodology, leading to optimal expressions
for operators and matrix elements.

A number of methodologies to calculate the angular parts of
matrix elements exists in literature.
Many existing codes for integrating the spin-angular parts of matrix elements
(Glass~\cite{Glass}, Glass and Hibbert~\cite{GH}, Grant~\cite{Grant},
Burke {\it et al }~\cite{BBD}) are
based on the computational scheme proposed by
Fano~\cite{Fano}. This methodology is based on having the total wavefunction of
an atom built from the antisymmetrized wavefunctions of separate shells, and
this antisymmetrization is done via coefficients of fractional parentage.
The shells are coupled one to another via their angular momenta. So, the
finding of matrix elements amounts to finding the recoupling matrices and the
coefficients of fractional parentage.

Suppose that we have a bra function with $u$ shells in $LS$ coupling:
\begin{equation}
\label{eq:ma-a}
\begin{array}[b]{c}
\psi _u^{bra}\left( LSM_LM_S\right) \equiv \\
\equiv (n_1l_1n_2l_2...n_ul_u\alpha _1L_1S_1\alpha
_2L_2S_2...\alpha _uL_uS_u{\cal A}LSM_LM_S|
\end{array}
\end{equation}
and a ket function:
\begin{equation}
\label{eq:ma-b}
\begin{array}[b]{c}
\psi _u^{ket}\left( L^{\prime }S^{\prime }M^{\prime}_LM^{\prime}_S\right) \equiv \\
\equiv |n_1l_1n_2l_2...n_ul_u\alpha _1^{\prime }L_1^{\prime }S_1^{\prime
}\alpha _2^{\prime }L_2^{\prime }S_2^{\prime
}...\alpha _u^{\prime }L_u^{\prime
}S_u^{\prime }{\cal A}^{\prime }L^{\prime }
S^{\prime}M^{\prime}_LM^{\prime}_S),
\end{array}
\end{equation}
where ${\cal A}$ stands for all intermediate quantum numbers, depending on the
order of coupling of momenta $L_iS_i$.
Using the Wigner-Eckart theorem in $LS$ space we shift from the
matrix element of any two-particle operator $\widehat{G}$
between functions (\ref{eq:ma-a}) and (\ref{eq:ma-b})
to the submatrix element (reduced matrix)

$(\psi _u^{bra}\left( LS\right) ||
\widehat{G}^{\left( \kappa _1\kappa _2k,\sigma_1\sigma _2k\right) }
||\psi _u^{ket}\left( L^{\prime
}S^{\prime }\right) )$ of this operator.
When the two-particle operator acts upon four distinct shells, then
the finding general expressions of
matrix elements, according to methodology by
Fano~\cite{Fano}, is based upon the formula

\begin{equation}
\label{eq:ma-c}
\begin{array}[b]{c}
(\psi _u^{bra}\left( LS\right) ||
\widehat{G}^{\left( \kappa _1\kappa _2k,\sigma_1\sigma _2k\right) }
||\psi _u^{ket}\left( L^{\prime }S^{\prime}\right) ) \sim \\
 \sim
\displaystyle {\sum_{n_{i} \lambda _{i} ,n_{j} \lambda _{j} ,
n_{i^{\prime }} \lambda _{i^{\prime}} ,n_{j^{\prime}} \lambda _{j^{\prime}} }}
\left( -1\right) ^\Delta
\left[ N_{i} \left( N_{j}-\delta(i,j) \right)
N_{i^{\prime}}^{\prime} \left( N_{j^{\prime}}^{\prime}-\delta(i^{\prime},j^{\prime})
\right) \right]^{1/2} \times \\
\times \left( l_{i}^{N_{i}}\;\alpha_{i} L_{i}S_{i}||l_{i}^{N_{i}-1}\;\left(
\alpha_{i}^{\prime } L^{\prime}_{i} S^{\prime }_{i} \right), l_{i} \right)
\left( l_{j}^{N_{j}}\;\alpha_{j} L_{j}S_{j}||l_{j}^{N_{j}-1}\;\left(
\alpha_{j}^{\prime } L^{\prime}_{j} S^{\prime }_{j} \right), l_{j} \right)
\times \\
\times
\left( l_{i^{\prime }}^{N_{i^{\prime }}}\;\alpha_{i^{\prime }}
L_{i^{\prime }}S_{i^{\prime }}||l_{i^{\prime }}^{N_{i^{\prime }}-1}\;\left(
\alpha_{i^{\prime }}^{\prime } L^{\prime}_{i^{\prime }}
S^{\prime }_{i^{\prime }} \right), l_{i^{\prime }} \right)
\left( l_{j^{\prime }}^{N_{j^{\prime }}}\;\alpha_{j^{\prime }}
L_{j^{\prime }}S_{j^{\prime }}||l_{j^{\prime }}^{N_{j^{\prime }}-1}\;\left(
\alpha_{j^{\prime }}^{\prime } L^{\prime}_{j^{\prime }}
S^{\prime }_{j^{\prime }} \right), l_{j^{\prime }} \right) \times \\
\times \left\{
R_{d} \left( \lambda _i,\lambda _j,\lambda _i^{\prime },\lambda _j^{\prime},
\kappa_1, \kappa_2, \kappa, \sigma_1, \sigma_2, \sigma, \Lambda ^{bra},
\Lambda ^{ket} \right) \right. \\
\times
\left( 1 + \delta(i,j) \delta(i^{\prime} ,j^{\prime} ) \right)^{-1}
\left( n_i\lambda _in_j\lambda _j||
g^{\left( \kappa _1\kappa _2k,\sigma_1\sigma _2k\right) }
||n_i^{\prime }\lambda _i^{\prime }n_j^{\prime }\lambda_j^{\prime }\right) - \\
- R_{e}  \left( \lambda _i,\lambda _j,\lambda _i^{\prime },\lambda _j^{\prime},
\kappa_1, \kappa_2, \kappa, \sigma_1, \sigma_2, \sigma, \Lambda ^{bra},
\Lambda ^{ket} \right) \\
\times  \left.
\left( 1 + \delta(i,j) )(1 - \delta(i^{\prime} ,j^{\prime} ) \right)
\left( n_i\lambda _in_j\lambda _j||
g^{\left( \kappa _1\kappa _2k,\sigma_1\sigma _2k\right) }
||n_j^{\prime }\lambda_j^{\prime }n_i^{\prime }\lambda _i^{\prime }\right)
\right\},
\end{array}
\end{equation}
where $\Delta$ is a phase factor, (see for example in \cite{Grant}),

$\Lambda ^{bra}\equiv \left( L_iS_i,L_jS_j,L_i^{\prime }S_i^{\prime
},L_j^{\prime }S_j^{\prime }\right) ^{bra}$ is the array for the bra function
shells' terms, and similarly for $\Lambda ^{ket}$.
The coefficient $\left( l^{N}\;\alpha LS||l^{N-1}\;\left(
\alpha^{\prime } L^{\prime} S^{\prime } \right) l \right)$
is a fractional parentage coefficient, and coefficients

$R_{d} \left( \lambda _i,\lambda _j,\lambda _i^{\prime },\lambda _j^{\prime},
\kappa_1, \kappa_2, \kappa, \sigma_1, \sigma_2, \sigma, \Lambda ^{bra},
\Lambda ^{ket} \right)$
and

$R_{e} \left( \lambda _i,\lambda _j,\lambda _i^{\prime },\lambda _j^{\prime},
\kappa_1, \kappa_2, \kappa, \sigma_1, \sigma_2, \sigma, \Lambda ^{bra},
\Lambda ^{ket} \right)$
are the recoupling matrices in $l$ and $s$ - spaces of direct and
exchange terms, respectively. For more detailes on recoupling matrices see in
Grant~\cite{Grant}, Burke {\it et al }~\cite{BBD}.

The summation in expression (\ref{eq:ma-c}) implies that the quantum numbers
$n$, $\lambda$ of all participating shells are included. There are four such
pairs of $n$, $\lambda$ in the sum.

In  essence, the Fano calculation scheme
consists of evaluating recoupling matrices. Although such an
approach uses  classical Racah algebra \cite{Racaha,Racahb,Racahc,Racahd}
on the level of coefficients of fractional parentage,
it may be necessary to
carry out multiple summations over intermediate terms. Due to
these summations and the complexity of the recoupling matrix
itself, the associated computer codes become rather time consuming.
Jucys and Vizbarait\. e~\cite{JV} proposed to use the two-electron coefficients
of fractioanl parentage instead of ordinary ones, in matrix elements'
calculations, but even that did not solve the abovementioned problems.
A solution to this problem was found by Burke {\it et al }~\cite{BBD}.
They tabulated
separate standard parts of recoupling matrices along with coefficients of
fractional parentage at the beginning of a calculation and used them further
on to calculate the needed coefficients. Computer codes by Glass~\cite{Glass},
Glass and Hibbert~\cite{GH}, Grant~\cite{Grant}, Burke {\it et al }~\cite{BBD}
utilize the
program NJSYM (Burke~\cite{Burke}) or NJGRAF (Bar-Shalom and
Klapisch~\cite{BSK}) for
the calculation of recoupling matrices. Both are rather time consuming
when calculating matrix elements of complex operators or
electronic configurations with many open shells.

In order to simplify the calculations, Cowan~\cite{Cowan} suggested that
matrix elements be grouped into ''Classes'' (see Cowan~\cite{Cowan}
Figure 13-5).
Unfortunately, this approach was not generalized to all two-electron operators.
Perhaps for this reason Cowan's approach is not very popular
although the program itself, based on this approach, is widely used.

Many approaches for the calculation of spin-angular coefficients
(Glass~\cite{Glass}, Glass and Hibbert~\cite{GH}, Grant~\cite{Grant},
Burke {\it et al }~\cite{BBD}) are based
on the usage of Racah algebra only on the level of coefficients of
fractional parentage. A few authors (Jucys and Savukynas~\cite{js},
Cowan~\cite{Cowan})
utilize the unit tensors, simplifying the calculations in this way, because
use can be made of the tables of unit tensors and selection rules
can be used prior to computation
to check whether the spin-angular coefficients
are zero or not. Moreover, the recoupling
matrices themselves have a simpler form. Unfortunately, these ideas were
applied only to diagonal matrix elements with respect to configurations,
though Cowan~\cite{Cowan} suggested the usage of unit tensors for
non-diagonal ones as well.

All the above mentioned approaches were applied in the coordinate
representation.
The second quantization formalism (Judd~\cite{J,jd},
Rudzikas and Kaniauskas~\cite{ra} and Rudzikas~\cite{r}) has a
number of advantages compared to coordinate representation. First of all, it
is much easier to find algebraic expressions for complex operators and
their matrix elements, when relying on second quantization formalism. It has
contributed significantly to the successful development of
perturbation theory (see Lindgren and Morrison~\cite{LM},
Merkelis {\it et al}~\cite{MGR},
and orthogonal operators (Uylings~\cite{U}), where three-particle
operators already occur. Uylings~\cite{Ub} suggested a fairly simple approach
for dealing with separate cases of three-particle operators.

Moreover, in the
second quantization approach the quasispin formalism was efficiently
developed by Innes~\cite{Innes}, \v Spakauskas {\it et al}~\cite{SKRa,SKRb},
Rudzikas and Kaniauskas~\cite{ra}, Fano and Rau~\cite{fr}.
The main advantage of this approach is
that applying the quasispin
method for calculating the matrix elements of any operator, we
can use all advantages of the new version of Racah algebra
(see Rudzikas~\cite{r}) for integration of spin-angular part of any one- and
two-particle operator. For example,
the reduced coefficients of fractional parentage are independent of the
occupation number of the shell
(see Rudzikas and Kaniauskas~\cite{ra}, Gaigalas {\it et al}~\cite{GRFb}).
All this enabled Merkelis and
Gaigalas~\cite{MG} to work out a general perturbation theory approach for
complex cases of several open shells. In the paper by Merkelis~\cite{MGraf}
a detailed review of a version of graphical methodology is presented
that allows one
to represent the operators graphically and to find the matrix
elements of these operators using diagrammatic technique.

The majority of methods and computer codes
of finding angular coefficients discussed above
were faced with a
number of problems, the main of these being:

\begin{itemize}
\item  The high demand of CPU time for calculating the angular parts of matrix
elements even on the modern computer. Therefore the high accuracy of
characteristics of atomic quantities somtimes is even  unattainable.

\item  The methods are applicable in practice only for the comparatively simple
systems, because in the programs based on classical Racah algebra the
treatment of recoupling matrices is rather complicated, especialy when
finding the matrix elements of Breit-Pauli operators between complex
configurations.

\item  The configurations with open $f$-shells must often be included in
theoretical calculations. This causes problems in a number of
methodologies, because the complete account of $f$- shells implies using
a large number of coefficients of fractional parentage.

\end{itemize}

Gaigalas and Rudzikas~\cite{GR}, Gaigalas {\it et al}~\cite{GRF},
Gaigalas {\it et al}~\cite{GBRF}, Gaigalas {\it et al}~\cite{GRFb}
and Gaigalas and Rudzikas~\cite{GRb}
suggested
an efficient and general approach for finding the spin-angular parts
of matrix elements of atomic interactions, relying on the combination of the
second-quantization approach in the coupled tensorial form, the generalized
graphical technique and angular momentum theory in orbital, spin and quasispin
spaces as well as on the symetry properties of the quantities considered.
This approach is free of previous shortcomings.

It is the main goal of this work to present all this methodology
consistently and in a unifying manner, paying the special attention to
its main ideas. Also, we aim at a detailed discussion of obtaining the
efficient tensorial expressions of a two-particle operator, as well as the
analytical expressions for recoupling matrices. In addition to that, in
this work we aim at comparing this methodology to other calculation
of angular coefficients schemes.

From what is said above we see that the treatment of angular parts of matrix
elements is a many-sided question.
The classical scheme by Fano was developed in various aspects. The methods were
developed that used Racah algebra  \cite{Racaha,Racahb,Racahc,Racahd}
on a higher level, or used
the new version of the Racah algebra (angular momentum theory in orbital, spin
and quasispin spaces) (see Rudzikas~\cite{rb}).
The ways were searched for, to obtain
the expressions that had simpler recoupling matrices. In order to discuss all
that in more detail, to compare the existing methods of obtaining the angular
parts and to mark the advantages of one method or another, the finding angular
parts must be looked at from different angles.

Thus, the first thing to
discuss is the expression for any physical operator (see sections 2,3,4),
because already the form of it determines the level of application of tensor
algebra in the calculation of matrix elements of this operator. Also, it is
very important to make clear to what extent the Racah algebra is exploited
in one or another methodology of angular parts treatment. This helps to mark
the advantages of one method against another (see section 5). In addition,
it is of importance to compare the ways of calculating the recoupling
matrix in various methods (see section 6).

\section{Tensorial Expressions for Two-particle Operators}

It is well known in the literature that a scalar two-particle operator may be
presented the following
tensorial form (see Jucys and Savukynas~\cite{js}, Glass~\cite{Glass}):

\begin{equation}
\label{eq:op-a}
\widehat{G}^{\left( \kappa_1 \kappa_2 k, \sigma_1 \sigma_2 k \right)}=
\displaystyle {\sum_{i > j}} g \left (r_{i},r_{j} \right)
\displaystyle {\sum_{p}} \left( -1 \right)^{k-p}
\left[ \widehat{g}_{i}^{\left( \kappa_1 \sigma_1 \right)} \times
\widehat{g}_{j}^{\left( \kappa_2 \sigma_2 \right)} \right]_{p-p}^{(k~k)} ,
\end{equation}
where $g \left (r_{i},r_{j} \right)$ is the radial part of operator,
$\widehat{g}_{i}^{\left( \kappa_1 \sigma_1 \right)}$ is a tensor acting
upon the orbital and spin variables of the $i$-th function,
$\kappa_1$, $\kappa_2$ are the ranks of operator acting in orbital space, and
$\sigma_1$, $\sigma_2$ are the ranks of operator acting in spin space.

All the above mentioned approaches were usually applied in the coordinate
representation. Now we will investigate into the second quantization formalism,
which is broadly applied in atomic physics as well.

A two-particle operator in second quantization method is written as follows:

\begin{equation}
\label{eq:op-b}
G^{\left( \kappa_1 \kappa_2 k, \sigma_1 \sigma_2 k \right)}=
\displaystyle
{\sum_{n_{i}l_{i},n_{j}l_{j},n_{i^{\prime }}l_{i^{\prime }},
n_{j^{\prime }} l_{j^{\prime }}}}
\widehat{G}(ij,i^{\prime} j^{\prime})=
\frac 12\displaystyle
{\sum_{i,j,i^{\prime },j^{\prime }}}a_ia_ja_{j^{\prime }}^{\dagger}
a_{i^{\prime
}}^{\dagger}\left( i,j|g|i^{\prime },j^{\prime }\right),
\end{equation}
where $i\equiv n_il_ism_{l_i}m_{s_i}$,
$\left( i,j|g|i^{\prime },j^{\prime }\right)$ is
the two-electron matrix element of operator
$G^{\left( \kappa_1 \kappa_2 k, \sigma_1 \sigma_2 k \right)}$, and
$a _{i}$ is the electron creation
and $a_{j}^{\dagger}$ electron annihilation operators.
Meanwhile two tensorial forms are well known in second quantization.
In the first form the operators of second quantization follow in the normal
order:

\begin{equation}
\label{eq:op-c}
\begin{array}[b]{c}
G_I^{\left( \kappa_1 \kappa_2 k, \sigma_1 \sigma_2 k \right)}=
\displaystyle {\sum_{n_il_i,n_jl_j,n_i^{\prime }l_i^{\prime },n_j^{\prime
}l_j^{\prime }}}\widehat{G}_I(ij,i^{\prime }j^{\prime })= \\ =-\frac 12
\displaystyle {\sum_{n_il_i,n_jl_j,n_i^{\prime }l_i^{\prime },n_j^{\prime
}l_j^{\prime }}}\displaystyle {\sum_{\kappa _{12}\kappa _{12}^{^{\prime
}}\sigma _{12}\sigma _{12}^{^{\prime }}}}\displaystyle {\sum_p}\left(
-1\right) ^{k-p}\left[ \kappa _{12},\kappa _{12}^{\prime },\sigma
_{12},\sigma _{12}^{\prime }\right] ^{1/2}\times \\ \times \left( n_i\lambda
_in_j\lambda _j||g^{\left( \kappa _1\kappa _2k,\sigma _1\sigma _2k\right)
}||n_i^{\prime }\lambda _i^{\prime }n_j^{\prime }\lambda _j^{\prime }\right)
\times \\
\times \left\{
\begin{array}{ccc}
l_i^{\prime } & l_j^{\prime } & \kappa _{12}^{\prime } \\
\kappa _1 & \kappa _2 & k \\
l_i & l_j & \kappa _{12}
\end{array}
\right\} \left\{
\begin{array}{ccc}
s & s & \sigma _{12}^{\prime } \\
\sigma _1 & \sigma _2 & k \\
s & s & \sigma _{12}
\end{array}
\right\} \times \\
\times \left[ \left[ a^{\left( \lambda _i\right) }\times a^{\left( \lambda
_j\right) }\right] ^{\left( \kappa _{12}\sigma _{12}\right) }\times \left[
\stackrel{\sim }{a}^{\left( \lambda _i^{\prime }\right) }\times \stackrel{%
\sim }{a}^{\left( \lambda _j^{\prime }\right) }\right] ^{\left( \kappa
_{12}^{\prime }\sigma _{12}^{\prime }\right) }\right] _{p-p}^{\left(
kk\right) },
\end{array}
\end{equation}
where
$\left[ a,b\right] =\left( 2a+1\right) \left( 2b+1\right) $,
$\lambda \equiv ls$,
$\left( n_i\lambda _in_j\lambda _j||g^{\left( \kappa _1\kappa
_2k,\sigma _1\sigma _2k\right) }||n_i^{\prime }\lambda _i^{\prime
}n_j^{\prime }\lambda _j^{\prime }\right) $ is the two-electron submatrix
(reduced matrix) element of operator
$G^{\left( \kappa_1 \kappa_2 \kappa, \sigma_1 \sigma_2 \sigma \right)}$
and tensor $\stackrel{\sim }{a}^{\left( \lambda \right) }$ is defined as
(see for example Rudzikas~\cite{r})

\begin{equation}
\label{eq:op-d}
\stackrel{\sim}{a}_{m_{ \lambda}}^{\left( \lambda \right)
}=\left( -1\right)^{ \lambda-m_{ \lambda}}a_{-m_{ \lambda}}
^{\dagger \left( \lambda \right)}.
\end{equation}

The product of tensors
$\left[ \left[ a^{\left( \lambda _i\right) }\times a^{\left( \lambda
_j\right) }\right] ^{\left( \kappa _{12}\sigma _{12}\right) }\times \left[
\stackrel{\sim }{a}^{\left( \lambda _i^{\prime }\right) }\times \stackrel{%
\sim }{a}^{\left( \lambda _j^{\prime }\right) }\right] ^{\left( \kappa
_{12}^{\prime }\sigma _{12}^{\prime }\right) }\right] _{p-p}^{\left(
kk\right) }$
denotes tensorial part of operator
$G_{I}^{\left( \kappa_1 \kappa_2 k, \sigma_1 \sigma_2 k \right)}$.

In another form the second quantization operators are coupled by pairs
consisting of electron creation and annihilation operators. In coupled
tensorial form:

\begin{equation}
\label{eq:op-e}
\begin{array}[b]{c}
G_{II}^{\left( \kappa_1 \kappa_2 \kappa, \sigma_1 \sigma_2 \sigma \right)}=
\displaystyle {\sum_{n_il_i,n_jl_j,n_i^{\prime }l_i^{\prime },n_j^{\prime
}l_j^{\prime }}}\widehat{G}_{II}(ij,i^{\prime }j^{\prime })= \\ =\frac 12
\displaystyle {\sum_{n_il_i,n_jl_j,n_i^{\prime }l_i^{\prime },n_j^{\prime
}l_j^{\prime }}}\displaystyle {\sum_p}\left( -1\right) ^{k-p}\left(
n_i\lambda _in_j\lambda _j||g^{\left( \kappa _1\kappa _2k,\sigma _1\sigma
_2k\right) }||n_i^{\prime }\lambda _i^{\prime }n_j^{\prime }\lambda
_j^{\prime }\right) \times \\
\times \left\{ \left[ \kappa _1,\kappa _2,\sigma
_1,\sigma _2\right] ^{-1/2}\times \left[ \left[ a^{\left( \lambda _i\right)
}\times
\stackrel{\sim }{a}^{\left( \lambda _i^{\prime }\right) }\right] ^{\left(
\kappa _1\sigma _1\right) }\times \left[ a^{\left( \lambda _j\right) }\times
\stackrel{\sim }{a}^{\left( \lambda _j^{\prime }\right) }\right] ^{\left(
\kappa _2\sigma _2\right) }\right] _{p-p}^{\left( kk\right) }- \right. \\
\left. -\left(
-1\right) ^{l_i+l_j^{\prime }}\left\{
\begin{array}{ccc}
\kappa _1 & \kappa _2 & k \\
l_j^{\prime } & l_i & l_j
\end{array}
\right\} \left\{
\begin{array}{ccc}
\sigma _1 & \sigma _2 & k \\
s & s & s
\end{array}
\right\} \left[ a^{\left( \lambda _i\right) }\times \stackrel{\sim }{a}%
^{\left( \lambda _j^{\prime }\right) }\right] _{p-p}^{\left( kk\right)
}\delta \left( n_jl_j,n_i^{\prime }l_i^{\prime }\right) \right\}.
\end{array}
\end{equation}

The expression (\ref{eq:op-c}) consists of only one tensorial product whereas
(\ref{eq:op-e}) has two, but the summation in the first formula is also over
intermediate ranks $\kappa _{12}$, $\sigma _{12}$, $\kappa _{12}^{\prime }$
and $\sigma _{12}^{\prime }$, complicating in this way the calculations. The
advantages or disadvantages of these alternative forms of arbitrary
two-electron operator may be revealed in practical applications.

In these forms the product of second quantization operators denotes
tensorial part of operator
$G^{\left( \kappa_1 \kappa_2 k, \sigma_1 \sigma_2 k \right)}$.
For instance, the tensorial structure of
electrostatic (Coulomb) electron interaction operator is
$\kappa _1=\kappa _2=k,\sigma _1=\sigma _2=0$
(Jucys and Savukynas~\cite{js}), and only the two-electron submatrix elements $%
\left( n_i\lambda _in_j\lambda _j||g^{\left( \kappa _1\kappa _2k,\sigma
_1\sigma _2k\right) }||n_i^{\prime }\lambda _i^{\prime }n_j^{\prime }\lambda
_j^{\prime }\right) $ of these operators are different. In the case of
electrostatic interaction:

\begin{equation}
\label{eq:op-f}
\begin{array}[b]{c}
\left( n_i\lambda _in_j\lambda _j||g_{Coulomb}^{\left( kk0,000\right)
}||n_i^{\prime }\lambda _i^{\prime }n_j^{\prime }\lambda _j^{\prime }\right)
= \\
=2\left[ k\right] ^{1/2}\left( l_i||C^{\left( k\right) }||l_i^{\prime
}\right) \left( l_j||C^{\left( k\right) }||l_j^{\prime }\right) R_k\left(
n_il_in_i^{\prime }l_i^{\prime },n_jl_jn_j^{\prime }l_j^{\prime }\right) .
\end{array}
\end{equation}

From (\ref{eq:op-f}), by (\ref{eq:op-c}) and (\ref{eq:op-e}), we finally obtain
the following two secondly quantized expressions for Coulomb operator
(see Merkelis {\it et al}~\cite{MKR}):

\begin{equation}
\label{eq:op-g}
\begin{array}[b]{c}
\widehat{V}_I=-\frac 12
\displaystyle {\sum_{n_il_in_jl_jn_i^{\prime }l_i^{\prime }n_j^{\prime
}l_j^{\prime }}}\displaystyle {\sum_{\kappa _{12}\sigma _{12}k}}\left(
-1\right) ^{l_j+l_i^{\prime }+k+\kappa _{12}}\left[ \kappa _{12},\sigma
_{12}\right] ^{1/2}\left( l_i||C^{\left( k\right) }||l_i^{\prime }\right)
\times \\ \times \left( l_j||C^{\left( k\right) }||l_j^{\prime }\right)
R_k\left( n_il_in_i^{\prime }l_i^{\prime },n_jl_jn_j^{\prime }l_j^{\prime
}\right) \left\{
\begin{array}{ccc}
l_i & l_i^{\prime } & k \\
l_j^{\prime } & l_i & \kappa _{12}
\end{array}
\right\} \times \\
\times \left[ \left[ a^{\left( \lambda _i\right) }\times a^{\left( \lambda
_j\right) }\right] ^{\left( \kappa _{12}\sigma _{12}\right) }\times \left[
\stackrel{\sim }{a}^{\left( \lambda _i^{\prime }\right) }\times \stackrel{%
\sim }{a}^{\left( \lambda _j^{\prime }\right) }\right] ^{\left( \kappa
_{12}\sigma _{12}\right) }\right] ^{\left( 00\right) },
\end{array}
\end{equation}
\begin{equation}
\label{eq:op-h}
\begin{array}[b]{c}
\widehat{V}_{II}=
\displaystyle {\sum_{n_il_in_jl_jn_i^{\prime }l_i^{\prime }n_j^{\prime
}l_j^{\prime }}}\displaystyle {\sum_k}\left( l_i||C^{\left( k\right)
}||l_i^{\prime }\right) \left( l_j||C^{\left( k\right) }||l_j^{\prime
}\right) R_k\left( n_il_in_i^{\prime }l_i^{\prime },n_jl_jn_j^{\prime
}l_j^{\prime }\right) \times \\
\times \left\{ \left[ k\right] ^{-1/2}\left[
\left[ a^{\left( \lambda _i\right) }\times
\stackrel{\sim }{a}^{\left( \lambda _i^{\prime }\right) }\right] ^{\left(
k0\right) }\times \left[ a^{\left( \lambda _j\right) }\times \stackrel{\sim
}{a}^{\left( \lambda _j^{\prime }\right) }\right] ^{\left( k0\right)
}\right] ^{\left( 00\right) }+ \right. \\
\left. +\left( 2\left[ l_i\right] \right)
^{-1/2}\left[ a^{\left( \lambda _i\right) }\times \stackrel{\sim }{a}%
^{\left( \lambda _j^{\prime }\right) }\right] ^{\left( 00\right) }\delta
\left( n_jl_j,n_i^{\prime }l_i^{\prime }\right) \right\} ,
\end{array}
\end{equation}

The tensorial expressions for orbit-orbit and other physical operators in
second quantization form may be obtained in the same manner.

It is worth mentioning that the expressions (\ref{eq:op-g}) and (\ref
{eq:op-h}) embrace, already in an operator form, both
the diagonal interaction terms, relative to configurations, and the non-diagonal ones.
Non-diagonal terms define the interaction between all the possible electron
distributions over the configurations considered, differing by quantum
numbers of not more than two electrons.

The merits of representing operators in one form or another (\ref{eq:op-g})
or (\ref{eq:op-h}) are mostly determined by the technique used to find
their matrix elements and quantities in terms of which they are expressed.

In the paper Gaigalas and Rudzikas ~\cite{GR} it was shown that the tensorial
forms (\ref{eq:op-a}), (\ref{eq:op-c}), (\ref{eq:op-e}) of two-particle
operator do not take full advantage of tensor algebra. The most characteristic
examples are when configurations considered have many open shells, or when
the non-diagonal matrix elements are seeked.

In the paper Gaigalas {\it et al}~\cite{GRF} the following optimal expression
of two-particle operator is proposed, which allows one to make the most
of the advantages of Racah algebra
(see Racah \cite{Racaha,Racahb,Racahc,Racahd}).

\begin{equation}
\label{eq:op-i}
\begin{array}[b]{c}
\widehat{G}^{\left( \kappa_1 \kappa_2 k, \sigma_1 \sigma_2 k \right)}\sim
\displaystyle {\sum_{\alpha}}
\displaystyle {\sum_{\kappa _{12},\sigma _{12},\kappa
_{12}^{\prime },\sigma _{12}^{\prime }}}\Theta \left( \Xi \right) \left\{
A_{p,-p}^{\left( kk\right) }\left( n_\alpha \lambda _\alpha ,\Xi \right)
\delta \left( u,1\right) \right. \\
+
\displaystyle {\sum_{\beta}}
\left[ B^{\left( \kappa _{12}\sigma
_{12}\right) }\left( n_\alpha \lambda _\alpha ,\Xi \right) \times C^{\left(
\kappa _{12}^{\prime }\sigma _{12}^{\prime }\right) }\left( n_\beta \lambda
_\beta ,\Xi \right) \right] _{p,-p}^{\left( kk\right) }\delta \left(
u,2\right) \\
+
\displaystyle {\sum_{\beta \gamma}}
\left[ \left[ D^{\left( l_\alpha s\right) }\times D^{\left( l_\beta
s\right) }\right] ^{\left( \kappa _{12}\sigma _{12}\right) }\times E^{\left(
\kappa _{12}^{\prime }\sigma _{12}^{\prime }\right) }\left( n_\gamma \lambda
_\gamma ,\Xi \right) \right] _{p,-p}^{\left( kk\right) }\delta \left(
u,3\right) \\
\left. +
\displaystyle {\sum_{\beta \gamma \delta}}
\left[ \left[ D^{\left( l_\alpha s\right) }\times D^{\left( l_\beta
s\right) }\right] ^{\left( \kappa _{12}\sigma _{12}\right) }\times \left[
D^{\left( l_\gamma s\right) }\times D^{\left( l_\delta s\right) }\right]
^{\left( \kappa _{12}^{\prime }\sigma _{12}^{\prime }\right) }\right]
_{p,-p}^{\left( kk\right) }\delta \left( u,4\right) \right\} .
\end{array}
\end{equation}

Whereas in traditional expressions, e. g. (\ref{eq:op-c}), the summation runs
over the principle and the orbital quantum numbers of open shells without
detailing these, in the expression written above the first term represents
the case of a two-particle operator acting upon the same shell
$n_\alpha \lambda _\alpha$, the second term corresponds to operator
$\widehat{G}^{\left( \kappa_1 \kappa_2 k, \sigma_1 \sigma_2 k \right)}$
acting upon two different shells
$n_\alpha \lambda _\alpha$, $n_\beta \lambda _\beta$. When operator
$\widehat{G}^{\left( \kappa_1 \kappa_2 k, \sigma_1 \sigma_2 k \right)}$
acts upon three shells the third term in
(\ref{eq:op-i}) must be considered and when it acts upon four - the
fourth one. We define in this expression the shells
$n_\alpha \lambda _\alpha$, $n_\beta \lambda _\beta$,
$n_\gamma \lambda _\gamma$, $n_\delta \lambda _\delta$
to be different.

The tensorial part of a two-particle operator is expressed in terms of
operators of the type
$A^{\left( kk\right) }\left( n\lambda ,\Xi \right)$,
$B^{\left( kk\right) }(n\lambda ,\Xi )$,
$C^{\left( kk\right) }(n\lambda ,\Xi )$,
$D^{\left( ls\right) }$,
$E^{\left( kk\right) }(n\lambda ,\Xi )$.
They correspond to one of the forms:

\begin{equation}
\label{eq:op-j}
a_{m_q}^{\left( q\lambda \right) },
\end{equation}

\begin{equation}
\label{eq:op-k}
\left[ a_{m_{q1}}^{\left( q\lambda \right) }\times
a_{m_{q2}}^{\left( q\lambda \right) }\right] ^{\left( \kappa _1\sigma
_1\right) },
\end{equation}

\begin{equation}
\label{eq:op-l}
\left[ a_{m_{q1}}^{\left( q\lambda \right) }\times \left[
a_{m_{q2}}^{\left( q\lambda \right) }\times a_{m_{q3}}^{\left( q\lambda
\right) }\right] ^{\left( \kappa _1\sigma _1\right) }\right] ^{\left( \kappa
_2\sigma _2\right) },
\end{equation}

\begin{equation}
\label{eq:op-m}
\left[ \left[ a_{m_{q1}}^{\left( q\lambda \right) }\times
a_{m_{q2}}^{\left( q\lambda \right) }\right] ^{\left( \kappa _1\sigma
_1\right) }\times a_{m_{q3}}^{\left( q\lambda \right) }\right] ^{\left(
\kappa _2\sigma _2\right) },
\end{equation}

\begin{equation}
\label{eq:op-n}
\left[ \left[ a_{m_{q1}}^{\left( q\lambda \right) }\times
a_{m_{q2}}^{\left( q\lambda \right) }\right] ^{\left( \kappa _1\sigma
_1\right) }\times \left[ a_{m_{q3}}^{\left( q\lambda \right) }\times
a_{m_{q4}}^{\left( q\lambda \right) }\right] ^{\left( \kappa _2\sigma
_2\right) }\right] ^{\left( kk\right) },
\end{equation}

For example, if we take a two-particle operator acting upon two shells,
then we see from expression (\ref{eq:op-i}) that the angular part of
two-particle operator is expressed via operators
$B^{\left( \kappa _{12}\sigma
_{12}\right) }\left( n_\alpha \lambda _\alpha ,\Xi \right)$ and
$C^{\left( \kappa _{12}^{\prime }\sigma _{12}^{\prime }\right) }\left( n_\beta
\lambda _\beta ,\Xi \right)$. In a case when the operator
$\widehat{G}^{\left( \kappa_1 \kappa_2 k, \sigma_1 \sigma_2 k \right)}$
acts in such a manner that two operators of second quantization act upon one
shell and two act upon another, the
$B^{\left( \kappa _{12}\sigma
_{12}\right) }\left( n_\alpha \lambda _\alpha ,\Xi \right)$ and
$C^{\left( \kappa _{12}^{\prime }\sigma _{12}^{\prime }\right) }\left( n_\beta
\lambda _\beta ,\Xi \right)$ are expressed as (\ref{eq:op-k}).
But in a case when three operators of second quantization act upon one shell
and one acts upon another, then
$B^{\left( \kappa _{12}\sigma
_{12}\right) }\left( n_\alpha \lambda _\alpha ,\Xi \right)$ and
$C^{\left( \kappa _{12}^{\prime }\sigma _{12}^{\prime }\right) }\left( n_\beta
\lambda _\beta ,\Xi \right)$ are expressed either as
(\ref{eq:op-j}) and
(\ref{eq:op-l}) or (\ref{eq:op-j}) and (\ref{eq:op-m}).

In writing down the expressions (\ref{eq:op-j}) - (\ref{eq:op-n})
the quasispin formalism was used, where
$a_{m_\lambda }^{\left(
\lambda \right) }$ and $\stackrel{\sim }{a}_{m_\lambda }^{\left( \lambda
\right) }$ are components of the tensor $a_{m_qm_\lambda }^{\left( q\lambda
\right) }$ , having in additional quasispin space the rank $q=\frac 12$ and
projections $m_q=\pm \frac 12$, i.e.

\begin{equation}
\label{eq:gimimas}
a_{\frac 12m_\lambda }^{\left(
q\lambda \right) }=a_{m_lm_s}^{\left( ls\right) }
\end{equation}

and

\begin{equation}
\label{eq:isnykimas}
a_{-\frac 12m_\lambda }^{\left( q\lambda \right) }=
\stackrel{\sim }{a}_{m_lm_s}^{\left( ls\right) }.
\end{equation}

In the expression (\ref{eq:op-i})
$u$ is the overall number of shells acted upon by a given
tensorial product of creation/annihilation operators. Parameter $\Xi $
implies the whole array of parameters (and sometimes an internal summation
over some of these is implied, as well) that connect the amplitudes $\Theta $
of tensorial products of creation/annihilation operators in the expression
(\ref{eq:op-i}) to these tensorial products (see Gaigalas {\it et al}~\cite{GRF}).
Also, attention must be paid to the fact that the ranks
$ \kappa_1$, $ \kappa_2$, $ \kappa$, $ \sigma_1 $, $ \sigma_2$ and $ \sigma $
are also included into the parameter $\Xi $.
The amplitudes $\Theta \left( \Xi \right) $ are all
proportional to the submatrix element of a two-particle operator $g$,

\begin{equation}
\label{eq:op-r}\Theta \left( \Xi \right) \sim \left( n_i\lambda _in_j\lambda
_j\left\| g\right\| n_{i^{\prime }}\lambda _{i^{\prime }}n_{j^{\prime
}}\lambda _{j^{\prime }}\right) .
\end{equation}

To obtain the expression of a concrete physical operator, analogous to
expression (\ref{eq:op-i}), the tensorial structure of the operator and the
two-particle matrix elements (\ref{eq:op-r}) must be known.
We shall investigate this now.

The {\it electrostatic (Coulomb) electron interaction} operator $H^{Coulomb}$
itself contains the tensorial structure

\begin{equation}
\label{eq:co-a}
\begin{array}[b]{c}
H^{Coulomb} \equiv
\displaystyle {\sum_{k}} H_{Coulomb}^{(kk0,000)}
\end{array}
\end{equation}
and its submatrix element is

\begin{equation}
\label{eq:co-b}
\begin{array}[b]{c}
\left( n_i\lambda _in_j\lambda _j\left\| H_{Coulomb}^{(kk0,000)}
\right\| n_{i^{\prime }}\lambda _{i^{\prime }}n_{j^{\prime }}\lambda
_{j^{\prime }}\right) = \\
2 [k]^{1/2}
\left( l_i\left\| C^{\left( k \right) }\right\| l_{i^{\prime }}\right)
\left( l_j\left\| C^{\left( k \right) }\right\| l_{j^{\prime }}\right)
R_{k}\left( n_il_i n_{i^{\prime }}l_{i^{\prime }},n_jl_j n_{j^{\prime
}}l_{j^{\prime }}\right).
\end{array}
\end{equation}

The {\it spin-spin} operator $H^{ss}$ itself contains tensorial structure
of two different types, summed over $k$ (Gaigalas and Rudzikas~\cite{GRb}),

\begin{equation}
\label{eq:ss-a}
\begin{array}[b]{c}
H^{ss} \equiv
\displaystyle {\sum_{k}}
\left[ H_{ss}^{(k+1 k-1 2,112)} + H_{ss}^{(k-1 k+1 2,112)} \right].
\end{array}
\end{equation}
Their submatrix elements are (Jucys and Savukynas~\cite{js})

\begin{equation}
\label{eq:ss-b}
\begin{array}[b]{c}
\left( n_i\lambda _in_j\lambda _j\left\| H_{ss}^{\left( k+1k-12,112\right)
}\right\| n_{i^{\prime }}\lambda _{i^{\prime }}n_{j^{\prime }}\lambda
_{j^{\prime }}\right) =\frac 3{
\sqrt{5}}\sqrt{\left( 2k+3\right) ^{\left( 5\right) }}\times  \\ \times
\left( l_i\left\| C^{\left( k+1\right) }\right\| l_{i^{\prime }}\right)
\left( l_j\left\| C^{\left( k-1\right) }\right\| l_{j^{\prime }}\right)
N^{k-1}\left( n_il_in_jl_j,n_{i^{\prime }}l_{i^{\prime }}n_{j^{\prime
}}l_{j^{\prime }}\right)
\end{array}
\end{equation}

\begin{equation}
\label{eq:ss-c}
\begin{array}[b]{c}
\left( n_i\lambda _in_j\lambda _j\left\| H_{ss}^{\left( k-1k+12,112\right)
}\right\| n_{i^{\prime }}\lambda _{i^{\prime }}n_{j^{\prime }}\lambda
_{j^{\prime }}\right) =\frac 3{
\sqrt{5}}\sqrt{\left( 2k+3\right) ^{\left( 5\right) }}\times  \\ \times
\left( l_i\left\| C^{\left( k-1\right) }\right\| l_{i^{\prime }}\right)
\left( l_j\left\| C^{\left( k+1\right) }\right\| l_{j^{\prime }}\right)
N^{k-1}\left( n_jl_jn_il_i,n_{j^{\prime }}l_{j^{\prime }}n_{i^{\prime
}}l_{i^{\prime }}\right),
\end{array}
\end{equation}
where we use a shorthand notation

$\left( 2k+3\right) ^{\left( 5\right) } \equiv \left( 2k+3\right)
\left( 2k+2\right)\left( 2k+1\right)\left( 2k\right)\left( 2k-1\right)$ and
radial integral (\ref{eq:ss-b}), (\ref{eq:ss-c}) is defined as in Glass and
Hibbert~\cite{GH}:

\begin{equation}
\label{eq:ss-d}
\begin{array}[b]{c}
N^k\left( n_il_in_jl_j,n_{i^{\prime }}l_{i^{\prime }}n_{j^{\prime
}}l_{j^{\prime }}\right) \\
=\frac{\alpha ^2}4\int_0^\infty \int_0^\infty P_i\left( r_1\right) P_j\left(
r_2\right) \frac{r_2^k}{r_1^{k+3}}\epsilon (r_1-r_2)P_{i^{\prime }}\left(
r_1\right) P_{j^{\prime }}\left( r_2\right) dr_1dr_{2,}
\end{array}
\end{equation}
where $\epsilon (x)$ is a Heaviside step-function,

\begin{equation}
\label{eq:ss-e}
\epsilon (x)=\left\{
\begin{array}{ll}
1 ; & \mbox{ for } x>0, \\ 0 ; & \mbox{ for } x\leq 0.
\end{array}
\right.
\end{equation}

The {\it spin-other-orbit} operator $H^{soo}$ itself contains tensorial
structure of six different types, summed over $k$
(see Gaigalas {\it et al}~\cite{GBRF}):
\begin{equation}
\label{eq:soo-a}
\begin{array}[b]{c}
H^{sso} \equiv
\displaystyle {\sum_{k}}
\left[ H_{sso}^{(k-1 k 1,101)} + H_{sso}^{(k-1 k 1,011)}+ \right. \\
\left. H_{sso}^{(k k 1,101)} + H_{sso}^{(k k 1,011)}+
H_{sso}^{(k+1 k 1,101)} + H_{sso}^{(k+1 k 1,011)}
\right].
\end{array}
\end{equation}
Their submatrix elements are:

\begin{equation}
\label{eq:soo-b}
\begin{array}[b]{c}
\left( n_i\lambda _in_j\lambda _j\left\| H_{soo}^{\left( k-1k1,\sigma
_1\sigma _21\right) }\right\| n_{i^{\prime }}\lambda _{i^{\prime
}}n_{j^{\prime }}\lambda _{j^{\prime }}\right) =2\cdot 2^{\sigma _2}\left\{
\left( 2k-1\right) \left( 2k+1\right) \right. \\
\times \left. \left( l_i+l_{i^{\prime }}-k+1\right) \left(
k-l_i+l_{i^{\prime }}\right) \left( k+l_i-l_{i^{\prime }}\right) \left(
k+l_i+l_{i^{\prime }}+1\right) \right\} ^{1/2} \\
\times \left( k\right) ^{-1/2}\left( l_i\left\| C^{\left( k\right) }\right\|
l_{i^{\prime }}\right) \left( l_j\left\| C^{\left( k\right) }\right\|
l_{j^{\prime }}\right) N^{k-2}\left( n_jl_jn_il_i,n_{j^{\prime
}}l_{j^{\prime }}n_{i^{\prime }}l_{i^{\prime }}\right),
\end{array}
\end{equation}

\begin{equation}
\label{eq:soo-c}
\begin{array}[b]{c}
\left( n_i\lambda _in_j\lambda _j\left\| H_{soo}^{\left( kk1,\sigma _1\sigma
_21\right) }\right\| n_{i^{\prime }}\lambda _{i^{\prime }}n_{j^{\prime
}}\lambda _{j^{\prime }}\right) =-2\cdot 2^{\sigma _2}\left( 2k+1\right)
^{1/2}\left( l_i\left\| C^{\left( k\right) }\right\| l_{i^{\prime }}\right)
\\
\times \left( l_j\left\| C^{\left( k\right) }\right\| l_{j^{\prime }}\right)
\left\{ \left( k\left( k+1\right) \right) ^{-1/2}\left( l_i\left(
l_i+1\right) -k\left( k+1\right) -l_{i^{\prime }}\left( l_{i^{\prime
}}+1\right) \right) \right. \\
\times \left\{ \left( k+1\right) N^{k-2}\left( n_jl_jn_il_i,n_{j^{\prime
}}l_{j^{\prime }}n_{i^{\prime }}l_{i^{\prime }}\right) -kN^k\left(
n_il_in_jl_j,n_{i^{\prime }}l_{i^{\prime }}n_{j^{\prime }}l_{j^{\prime
}}\right) \right\} \\
\left. -2\left( k\left( k+1\right) \right) ^{1/2}V^{k-1}\left(
n_il_in_jl_j,n_{i^{\prime }}l_{i^{\prime }}n_{j^{\prime }}l_{j^{\prime
}}\right) \right\},
\end{array}
\end{equation}

\begin{equation}
\label{eq:soo-d}
\begin{array}[b]{c}
\left( n_i\lambda _in_j\lambda _j\left\| H_{soo}^{\left( k+1k1,\sigma
_1\sigma _21\right) }\right\| n_{i^{\prime }}\lambda _{i^{\prime
}}n_{j^{\prime }}\lambda _{j^{\prime }}\right) =2\cdot 2^{\sigma _2}\left\{
\left( 2k+1\right) \left( 2k+3\right) \right. \\
\times \left. \left( l_i+l_{i^{\prime }}-k\right) \left( k-l_i+l_{i^{\prime
}}+1\right) \left( k+l_i-l_{i^{\prime }}+1\right) \left( k+l_i+l_{i^{\prime
}}+2\right) \right\} ^{1/2} \\
\times \left( k+1\right) ^{-1/2}\left( l_i\left\| C^{\left( k\right)
}\right\| l_{i^{\prime }}\right) \left( l_j\left\| C^{\left( k\right)
}\right\| l_{j^{\prime }}\right) N^k\left( n_il_in_jl_j,n_{i^{\prime
}}l_{i^{\prime }}n_{j^{\prime }}l_{j^{\prime }}\right) .
\end{array}
\end{equation}
The radial integrals in  (\ref{eq:soo-b}) - (\ref{eq:soo-d}) are (see
Glass and Hibbert~\cite{GH}):

\begin{equation}
\label{eq:soo-e}
\begin{array}[b]{c}
V^k\left( n_il_in_jl_j,n_{i^{\prime }}l_{i^{\prime }}n_{j^{\prime
}}l_{j^{\prime }}\right) \\
=\frac{\alpha ^2}4\int_0^\infty \int_0^\infty P_i\left( r_1\right) P_j\left(
r_2\right) \frac{r_{<}^{k-1}}{r_{>}^{k+2}}r_2\frac \partial {\partial
r_1}P_{i^{\prime }}\left( r_1\right) P_{j^{\prime }}\left( r_2\right)
dr_1dr_2.
\end{array}
\end{equation}

The tensorial form of {\it orbit-orbit}  operator is
(see Eissner {\it et al}~\cite{EJN})

\begin{equation}
\label{eq:oo-a}H^{oo}=\displaystyle
\sum_k\left(
H_{oo1}^{(kk0,000)}+H_{oo2}^{(kk0,000)}+H_{oo3}^{(kk0,000)}+H_{oo4}^{(kk0,00)}\right) ,
\end{equation}

The sum of submatrix elements of three terms
$H_{oo1}^{(kk0,000)}$, $H_{oo2}^{(kk0,000)}$ and
$H_{oo4}^{(kk0,000)}$ is equal to
(see Badnell~\cite{Badnell}):

\begin{equation}
\label{eq:oo-b}
\begin{array}[b]{c}
\left( n_i\lambda _in_j\lambda _j\left\|
H_{oo1}^{(kk0,000)} + H_{oo2}^{(kk0,000)} +
H_{oo4}^{(kk0,000)}
\right\| n_{i^{\prime }}\lambda _{i^{\prime
}}n_{j^{\prime }}\lambda _{j^{\prime }}\right) =   \\
= \left\{ -2k\left( k+1\right) \left( T^{k+1}\left(
n_il_in_jl_j,n_{i^{\prime }}l_{i^{\prime }}n_{j^{\prime }}l_{j^{\prime
}}\right) -T^{k-1}\left( n_il_in_jl_j,n_{i^{\prime }}l_{i^{\prime
}}n_{j^{\prime }}l_{j^{\prime }}\right) \right) \right. \\
+\left( l_i\left( l_i+1\right) -k\left( k+1\right) -l_{i^{\prime }}\left(
l_{i^{\prime }}+1\right) \right)
\\ \times
\left( U^{k+1}\left(
n_il_in_jl_j,n_{i^{\prime }}l_{i^{\prime }}n_{j^{\prime }}l_{j^{\prime
}}\right) -U^{k-1}\left( n_il_in_jl_j,n_{i^{\prime }}l_{i^{\prime
}}n_{j^{\prime }}l_{j^{\prime }}\right) \right)  \\
+\left( l_j\left( l_j+1\right) -k\left( k+1\right) -l_{j^{\prime }}\left(
l_{j^{\prime }}+1\right) \right)
\\ \times
\left( U^{k+1}\left(
n_jl_jn_il_i,n_{j^{\prime }}l_{j^{\prime }}n_{i^{\prime }}l_{i^{\prime
}}\right) -U^{k-1}\left( n_jl_jn_il_i,n_{j^{\prime }}l_{j^{\prime
}}n_{i^{\prime }}l_{i^{\prime }}\right) \right)   \\
+\frac 12\left( l_i\left( l_i+1\right) -k\left( k+1\right) -l_{i^{\prime
}}\left( l_{i^{\prime }}+1\right) \right) \left( l_j\left( l_j+1\right)
-k\left( k+1\right) -l_{j^{\prime }}\left( l_{j^{\prime }}+1\right) \right)
 \\
\times \left[
\frac{k-2}{k\left( 2k-1\right) }\left( N^{k-2}\left(
n_il_in_jl_j,n_{i^{\prime }}l_{i^{\prime }}n_{j^{\prime }}l_{j^{\prime
}}\right) +N^{k-2}\left( n_jl_jn_il_i,n_{j^{\prime }}l_{j^{\prime
}}n_{i^{\prime }}l_{i^{\prime }}\right) \right) \right.   \\
\left. \left. -\frac{k+3}{\left( k+1\right) \left( 2k+3\right) }\left( N^k\left(
n_il_in_jl_j,n_{i^{\prime }}l_{i^{\prime }}n_{j^{\prime }}l_{j^{\prime
}}\right) +N^k\left( n_jl_jn_il_i,n_{j^{\prime }}l_{j^{\prime }}n_{i^{\prime
}}l_{i^{\prime }}\right) \right) \right] \right\}
\\ \times
\left( 1-\delta \left( k,0\right) \right),
\end{array}
\end{equation}
where
\begin{equation}
\label{eq:oo-c}
\begin{array}[b]{c}
T^k\left( n_il_in_jl_j,n_{i^{\prime }}l_{i^{\prime }}n_{j^{\prime
}}l_{j^{\prime }}\right) =\frac{\alpha ^2}{4\left( 2k+1\right) }
\\ \times
\int_0^\infty \int_0^\infty P_i\left( r_1\right) P_j\left(
r_2\right) \frac{r_<^k}{r_>^{k+1}}
\left(\frac{\partial}{\partial r_1}+\frac{1}{r_1}\right)
P_{i^{\prime }}\left( r_1\right)
\left(\frac{\partial}{\partial r_2}+\frac{1}{r_2}\right)
P_{j^{\prime }}\left( r_2\right) dr_1dr_{2},
\end{array}
\end{equation}

\begin{equation}
\label{eq:oo-d}
\begin{array}[b]{c}
U^k\left( n_il_in_jl_j,n_{i^{\prime }}l_{i^{\prime }}n_{j^{\prime
}}l_{j^{\prime }}\right) =\frac{\alpha ^2}{4\left( 2k+1\right) }
\\ \times
\int_0^\infty \int_0^\infty P_i\left( r_1\right) P_j\left(
r_2\right)\left(
(k-1)\frac{r_2^k}{r_1^{k+2}}\epsilon (r_1-r_2)
-(k+2)\frac{r_1^{k-1}}{r_2^{k+1}}\epsilon (r_2-r_1)\right)
\\ \times
P_{i^{\prime }}\left( r_1\right)
\left(\frac{\partial}{\partial r_2}+\frac{1}{r_2}\right)
P_{j^{\prime }}\left( r_2\right) dr_1dr_{2}.
\end{array}
\end{equation}

The submatrix element of remaining term $H_{oo3}^{(kk0,000)}$ is:

\begin{equation}
\label{eq:oo-e}
\begin{array}[b]{c}
\left( n_i\lambda _in_j\lambda _j\left\|H_{oo3}^{(kk0,000)}
\right\| n_{i^{\prime }}\lambda _{i^{\prime
}}n_{j^{\prime }}\lambda _{j^{\prime }}\right) = 2
\sqrt{2k+1}\frac 1{k(k+1)}  \\ \times
\left( \left( l_i+l_{i^{\prime
}}+k+2\right) \left( l_i+l_{i^{\prime }}-k\right) \left( l_i-l_{i^{\prime
}}+k+1\right) \left( l_{i^{\prime }}-l_i+k+1\right) \right.
 \\  \times
\left. \left( l_j+l_{j^{\prime }}+k+2\right)
\times \left( l_j+l_{j^{\prime }}-k\right) \left( l_j-l_{j^{\prime
}}+k+1\right) \left( l_{j^{\prime }}-l_j+k+1\right) \right) ^{1/2}
 \\  \times
\left( l_i||C^{\left( k+1\right) }||l_{i^{\prime }}\right) \left(
l_j||C^{\left( k+1\right) }||l_{j^{\prime }}\right)
 \\ \times
\left( N^{k-1}\left(
n_il_in_jl_j,n_{i^{\prime }}l_{i^{\prime }}n_{j^{\prime }}l_{j^{\prime
}}\right) +N^{k-1}\left( n_jl_jn_il_i,n_{j^{\prime }}l_{j^{\prime
}}n_{i^{\prime }}l_{i^{\prime }}\right) \right) .
\end{array}
\end{equation}

The rest of two-particle Breit-Pauli operators that we did not investigate
so far are the two-body Darwin and spin-spin-contact terms.
They do not bring any additional difficulties into the investigation of
Hamiltonian, but for the sake of completeness of presentation we will
discuss them briefly

The {\it  two-body Darwin} operator $H_{D2}$ (see for more detail
Nikitin and Rudzikas~\cite{NR}), as well as the
{\it spin-spin-contact} operator $H_{SSC}$
(see Shalit and Talmi~\cite{Shalit} and Feneuille~\cite{Feneuille}),
both have the following tensorial structure:

\begin{equation}
\label{eq:oo-f}
\begin{array}[b]{c}
H \equiv
\displaystyle {\sum_{k}} H^{(kk0,000)}.
\end{array}
\end{equation}
These two terms are included into calculation by adding
to the radial integral
$R_{k}\left( n_il_i n_{i^{\prime }}l_{i^{\prime }},n_jl_j n_{j^{\prime
}}l_{j^{\prime }}\right)$ a term

\begin{eqnarray*}
\label{eq:oo-g}
\left( 2k+1 \right)
X\left( n_il_i n_{i^{\prime }}l_{i^{\prime }},n_jl_j n_{j^{\prime
}}l_{j^{\prime }}\right),
\end{eqnarray*}
where

\begin{equation}
\label{eq:oo-h}
\begin{array}[b]{c}
X\left( n_il_in_jl_j,n_{i^{\prime }}l_{i^{\prime }}n_{j^{\prime
}}l_{j^{\prime }}\right) \\
=\frac{\alpha ^2}4\int_0^\infty \int_0^\infty P_i\left( r_1\right) P_j\left(
r_2\right) \frac{1}{r_1^{2}} \delta \left( r_1-r_2 \right)
P_{i^{\prime }}\left(
r_1\right) P_{j^{\prime }}\left( r_2\right) dr_1dr_{2}.
\end{array}
\end{equation}
%GGG

The expression (\ref{eq:op-i}) has a series of terms, and thus at a first
glance seems to be difficult to apply. For this purpose in the next sections
we shall discuss in more detail:
\begin{itemize}
\item Obtaining the expression by Wick's theorem.
\item The compact written form of all terms, using the extended graphical
technique.
\item Obtaining the values of recoupling matrix and of the standard quantities.
We shall also compare the existing methodologies of finding angular parts,
showing the advantages and shortcomings of one methodology or another.
\end{itemize}

This methodology is fully applicable also for the one-particle operator
(see Gaigalas and Rudzikas~\cite{GR}, Gaigalas {\it et al}~\cite{GRF}).
As this operator does not cause big problems in the atomic physics,
we will not stop for its details.

\section{ Wick's theorem }

Wick's theorem in the second quantization formalism is formulated as follows
(see Wick~\cite{W}; Bogoliubov and Shirkov~\cite{BS}):
If $A$ is a product of creation and annihilation operators, then

\begin{equation}
\label{eq:v-aa}
\begin{array}{c}
A = \left\{ A \right\} + \left\{ \overline{A} \right\},
\end{array}
\end{equation}
where $\left\{ A \right\}$ represents the normal form of $A$ and
$\left\{ \overline{A} \right\}$ represents the sum  of the normal-ordered
terms obtained by making all possible single, double, ...
contractions within $A$. Based on the identification in
Bogoliubov and Shirkov~\cite{BS}, the operator is presented in normal form
when all of the operators of annihilation included in it are transferred to
the right of the creation operators.

Usually, Wick's theorem is applied when treating complex operators that are
represented by a large number of second quantization operators in a
non-normal product form. In atomic physics such operators are used in
perturbation theory
(see Fetter and Wale\v cka~\cite{FW}; Lindgren and Morrison~\cite{LM};
Merkelis {\it et al}~\cite{MGR})
and in the orthogonal operator method (see Uylings~\cite{U}).
Most often the Wick's theorem is applied to the products of second quantization
operators that are not tensorially coupled
(see Lindgren and Morrison~\cite{LM}). While applying the perturbation
theory in an extended model space, two different groups of second quantization
operators are defined (see for details in Lindgren and Morrison~\cite{LM}).
The second quantization operators acting upon core states
belong to one group, whereas the operators acting upon open and excited shells
belong to another one. These two groups are very different in applying Wick's
theorem to them.

In the first group, the creation operators are re-named to annihilation
operators and are called the hole absorption operators, while the annihilation
operators are re-named to creation operators and are called the hole
creation operators. The creation operators of the second group are called the
particle creation operators, while annihilation ones are called the particle
absorption
operators. Such a division of second quantization operators into two groups
is called the particle-hole formalism.

Merkelis {\it et al}~\cite{MGR} have proposed to use the so-called graphical
analogue of Wick's theorem in perturbation theory
(see Gaigalas {\it et al}~\cite{gga}, Gaigalas~\cite{gg}). It is an
efficient tool for
obtaining the normal products of second quantization operators in a coupled
tensorial form. Before applying this theorem, particular second quantization
terms are in a normal product in coupled form. In addition, this theorem
is applied in the particle-hole formalism, too
(see Merkelis {\it et al}~\cite{MGR}, Merkelis {\it et al}~\cite{MGKR}).

In all the cases mentioned above, the Wick's theorem is applied for the
most general case of operators, i.e. the shells that are acted upon are
not detailed. But, in the case of the extended model space, the group that the
second quantization operators belong to, depending on the electronic structure
of atom or ion under investigation, is defined.

Wick's theorem is not applied in investigations of ordinary physical operators.
Gaigalas {\it et al}~\cite{GRF} proposed a new version of the Wick's theorem
application, where the optimal tensorial expression of any two-particle
operator is easily obtained. The specificity of Wick's theorem application
in this case lies in applying it only when the shells that are acted upon by
the secondary quantization operators are known, i.e. it is applied for each
term $\widehat{G}(ij,i'j')$ separately. Such an interpretation of Wick's
theorem bears similarity with the particle-hole formalism. The only difference
is that in this case the second quantization operators are differentiated
formally not on the basis of structure of atom under investigation, but on
the basis of shells acted upon.

This is done in the following way. The second quantization operators acting
upon a shell with a lowest index are attributed to the first group. Those
acting upon a shell with a next-lowest index are attributed to the second
group, etc. In the most general case we have four distinct groups.

For example, suppose we have a two-particle operator $\widehat{G}(25,32)$,

\begin{equation}
\label{eq:v-a}
\begin{array}{c}
\widehat{G}(25,32) = \frac{1}{2}~a_2a_5a^{\dagger}_2a^{\dagger}_3~~(2,5~|g|3,2~),
\end{array}
\end{equation}
(where $2\equiv n_2l_2 sm_{l_2}m_{s_2}$, $3\equiv n_3l_3 sm_{l_3}m_{s_3}$,
$5\equiv n_5l_5 sm_{l_5}m_{s_5}$), the matrix element of which between the
functions

$\psi _5^{bra}\left( LS M_LM_S \right) \equiv (n_1l_1 n_2l_2 L_{12}S_{12}
n_3l_3 L_{123}S_{123} n_4 l_4 L_{1234}S_{1234} n_5 l_5 LS|$
 and
$\psi _5^{ket}\left( L'S' M'_LM'_S \right) \equiv |n_1l_1 n_2l_2 L'_{12}S'_{12}
n_3l_3 L'_{123}S'_{123} n_4 l_4 L'_{1234}S'_{1234} n_5 l_5 L'S')$
we must obtain. Then the operators acting upon the second shell are
attributed to the first group, the ones acting upon the third shell - to the
second, and the ones acting upon the fifth - to the third group.

Assuming that all the operators from first group are the creation ones, and
the rest are annihilation operators, we apply the Wick's theorem. Thus we
obtain that the operators from the first group are one beside another, and
all are positioned after the operators from the first group. Thus we obtain:

\begin{equation}
\label{eq:v-b}
\begin{array}{c}
-a_2a^{\dagger}_2a_5a^{\dagger}_3.
\end{array}
\end{equation}

After that, we apply Wick's theorem  assuming that the operators from
the first and the second groups are creation ones, and the  rest of them are
annihilation operators. Thus we obtain that the operators from the second
group are one beside another, and all are positioned after the first
group operators:

\begin{equation}
\label{eq:v-c}
\begin{array}{c}
a_2a^{\dagger}_2a^{\dagger}_3a_5.
\end{array}
\end{equation}

If in the product that we investigate there are operators of second
quantization acting upon four distinct shells, then we apply Wick's theorem
once again, assuming that operators from the first, second and third groups
are creation ones, and from the fourth group - annihilation ones.
In this case the Wick's theorem is applied to the second quantization operators
in uncoupled form.

From (\ref{eq:op-b}) we see that in second quantization a two-particle operator
is written as a sum, where parameters $i$, $j$, $i'$, $j'$ run over all
possible arrays of quantum numbers. So, the greater the number of open shells
in bra and ket functions, the greater the number of terms
$\widehat{G}(ij,i'j')$ in the expression of two-particle operator. It is
obvious that all these terms must be systematized in order to obtain in general
case the most efficient tensorial expression of a two-particle operator, in
the way described above.

%........................................................................
\clearpage
\begin{table}
\begin{center}
\caption{Distributions of shells, upon which the second quantization
operators are acting, that appear in the submatrix elements of any
two-particle operator, when bra and ket functions have $u$ open
shells (see Gaigalas {\it et al}~\cite{GRF})}
\label{pasis}
  \begin{tabular}{|r|c|c|c|c|c|} \hline
     &  &  &  &  & \\
    {\bf No.} & {\bf $a_{i}$ } & {\bf $a_{j}$ } & {\bf $a_{i'}^{\dagger}$ } &
    {\bf $a_{j'}^{\dagger}$ }
& {\bf submatrix element} \\
     &  &  &  &  & \\ \hline \hline
{\bf  1.} & $\alpha$  & $\alpha$ &  $\alpha$ & $\alpha$
& $ (...n_{\alpha}l_{\alpha}^{N_{\alpha}}...||
\widehat{G}(n_il_in_jl_jn_i^{\prime}l_i^{\prime }n_j^{\prime}l_j^{\prime })
||...n_{\alpha}l_{\alpha}^{N_{\alpha}}... )$ \\  \hline  \hline
{\bf  2.} & $\alpha$  & $\beta$  &  $\alpha$ &  $\beta$  & \\
{\bf  3.} & $\beta$   & $\alpha$ & $\beta$  & $\alpha$
& $ (...n_{\alpha}l_{\alpha}^{N_{\alpha}}...n_{\beta}l_{\beta}^{N_{\beta}}...||$ \\
{\bf  4.} & $\alpha$  & $\beta$  & $\beta$  & $\alpha$ &
$\widehat{G}
(n_il_in_jl_jn_i^{\prime}l_i^{\prime }n_j^{\prime}l_j^{\prime })$ \\
{\bf  5.} & $\beta$   & $\alpha$ & $\alpha$ & $\beta$
&$||...n_{\alpha}l_{\alpha}^{N_{\alpha}}...n_{\beta}l_{\beta}^{N_{\beta}}...)$ \\    \hline
{\bf  6.} & $\alpha$ & $\alpha$  & $\beta$  & $\beta$
& $ (...n_{\alpha}l_{\alpha}^{N_{\alpha}}...n_{\beta}l_{\beta}^{N_{\beta}}...||
\widehat{G}
||...n_{\alpha}l_{\alpha}^{N_{\alpha}-2}...n_{\beta}l_{\beta}^{N_{\beta}+2}...)$ \\    \hline
{\bf  7.} & $\beta$    & $\alpha$ & $\alpha$ & $\alpha$  & \\
{\bf  8.} & $\alpha$   & $\beta$  & $\alpha$ & $\alpha$
& $(...n_{\alpha}l_{\alpha}^{N_{\alpha}}...n_{\beta}l_{\beta}^{N_{\beta}}...||$ \\
{\bf  9.} & $\beta$    & $\beta$  & $\beta$  & $\alpha$  &
$\widehat{G}
(n_il_in_jl_jn_i^{\prime}l_i^{\prime }n_j^{\prime}l_j^{\prime })$ \\
{\bf 10.} & $\beta$   & $\beta$  & $\alpha$ & $\beta$
& $||...n_{\alpha}l_{\alpha}^{N_{\alpha}+1}...n_{\beta}l_{\beta}^{N_{\beta}-1}...)$ \\    \hline
 \hline
{\bf 11.} & $\beta$   & $\gamma$ & $\alpha$ & $\gamma$  & \\
{\bf 12.} & $\gamma$  & $\beta$  & $\gamma$ & $\alpha$
& $ (...n_{\alpha}l_{\alpha}^{N_{\alpha}}n_{\beta}l_{\beta}^{N_{\beta}}
n_{\gamma}l_{\gamma}^{N_{\gamma}}...||$ \\
{\bf 13.} & $\gamma$  & $\beta$  & $\alpha$ & $\gamma$  &
$\widehat{G}
(n_il_in_jl_jn_i^{\prime}l_i^{\prime }n_j^{\prime}l_j^{\prime })$  \\
{\bf 14.} & $\beta$   & $\gamma$ & $\gamma$ & $\alpha$
& $||...n_{\alpha}l_{\alpha}^{N_{\alpha}+1}n_{\beta}l_{\beta}^{N_{\beta}-1}
n_{\gamma}l_{\gamma}^{N_{\gamma}}...)$ \\    \hline
{\bf 15.} & $\gamma$  & $\gamma$ & $\alpha$ & $\beta$
& $ (...n_{\alpha}l_{\alpha}^{N_{\alpha}}n_{\beta}l_{\beta}^{N_{\beta}}
n_{\gamma}l_{\gamma}^{N_{\gamma}}...||
\widehat{G}
(n_il_in_jl_jn_i^{\prime}l_i^{\prime }n_j^{\prime}l_j^{\prime })$ \\
{\bf 16.} & $\gamma$  & $\gamma$ & $\beta$  & $\alpha$
& $||...n_{\alpha}l_{\alpha}^{N_{\alpha}+1}n_{\beta}l_{\beta}^{N_{\beta}+1}
n_{\gamma}l_{\gamma}^{N_{\gamma}-2}...)$ \\    \hline
{\bf 17.} & $\alpha$  & $\beta$  & $\gamma$ & $\gamma$
& $ (...n_{\alpha}l_{\alpha}^{N_{\alpha}}n_{\beta}l_{\beta}^{N_{\beta}}
n_{\gamma}l_{\gamma}^{N_{\gamma}}...||
\widehat{G}
(n_il_in_jl_jn_i^{\prime}l_i^{\prime }n_j^{\prime}l_j^{\prime })$ \\
{\bf 18.} & $\beta$   & $\alpha$ & $\gamma$ & $\gamma$
& $||...n_{\alpha}l_{\alpha}^{N_{\alpha}-1}n_{\beta}l_{\beta}^{N_{\beta}-1}
n_{\gamma}l_{\gamma}^{N_{\gamma}+2}...)$ \\    \hline
 \hline
{\bf 19.} & $\alpha$  & $\beta$  & $\gamma$ & $\delta$  & \\
{\bf 20.} & $\beta$   & $\alpha$ & $\gamma$ & $\delta$
& $ (n_{\alpha}l_{\alpha}^{N_{\alpha}}n_{\beta}l_{\beta}^{N_{\beta}}
n_{\gamma}l_{\gamma}^{N_{\gamma}}n_{\delta}l_{\delta}^{N_{\delta}}||$ \\
{\bf 21.} & $\alpha$  & $\beta$  & $\delta$ & $\gamma$  &
$\widehat{G}
(n_il_in_jl_jn_i^{\prime}l_i^{\prime }n_j^{\prime}l_j^{\prime })$ \\
{\bf 22.} & $\beta$   & $\alpha$ & $\delta$ & $\gamma$
& $||n_{\alpha}l_{\alpha}^{N_{\alpha}-1}n_{\beta}l_{\beta}^{N_{\beta}-1}
n_{\gamma}l_{\gamma}^{N_{\gamma}+1}n_{\delta}l_{\delta}^{N_{\delta}+1})$ \\    \hline
 \end{tabular}
\end{center}
\end{table}

\clearpage
\vspace {0.5in} Table 1 (continued) \vspace {0.1in}

\begin{center}
%\caption{Table 1 (continued)}
\begin{tabular}{|l|l|l|l|l|c|} \hline
     &  &  &  &  & \\
    {\bf No.} & {\bf $a_{i}$ } & {\bf $a_{j}$ } & {\bf $a_{i'}^{\dagger}$ }
     & {\bf $a_{j'}^{\dagger}$ }
& {\bf submatrix element} \\
     &  &  &  &  & \\ \hline \hline
{\bf 23.} & $\gamma$  & $\delta$ & $\alpha$ & $\beta$  & \\
{\bf 24.} & $\gamma$  & $\delta$ & $\beta$  & $\alpha$
& $(n_{\alpha}l_{\alpha}^{N_{\alpha}}n_{\beta}l_{\beta}^{N_{\beta}}
n_{\gamma}l_{\gamma}^{N_{\gamma}}n_{\delta}l_{\delta}^{N_{\delta}}||$ \\
{\bf 25.} & $\delta$  & $\gamma$ & $\alpha$ & $\beta$  &
$\widehat{G}
(n_il_in_jl_jn_i^{\prime}l_i^{\prime }n_j^{\prime}l_j^{\prime })$ \\
{\bf 26.} & $\delta$  & $\gamma$ & $\beta$  & $\alpha$
& $||n_{\alpha}l_{\alpha}^{N_{\alpha}+1}n_{\beta}l_{\beta}^{N_{\beta}+1}
n_{\gamma}l_{\gamma}^{N_{\gamma}-1}n_{\delta}l_{\delta}^{N_{\delta}-1})$ \\    \hline
{\bf 27.} & $\alpha$  & $\gamma$ & $\beta$  & $\delta$  & \\
{\bf 28.} & $\alpha$  & $\gamma$ & $\delta$ & $\beta$
& $ (n_{\alpha}l_{\alpha}^{N_{\alpha}}n_{\beta}l_{\beta}^{N_{\beta}}
n_{\gamma}l_{\gamma}^{N_{\gamma}}n_{\delta}l_{\delta}^{N_{\delta}}||$ \\
{\bf 29.} & $\gamma$  & $\alpha$ & $\delta$ & $\beta$  &
$\widehat{G}
(n_il_in_jl_jn_i^{\prime}l_i^{\prime }n_j^{\prime}l_j^{\prime })$ \\
{\bf 30.} & $\gamma$  & $\alpha$ & $\beta$  & $\delta$
& $||n_{\alpha}l_{\alpha}^{N_{\alpha}-1}n_{\beta}l_{\beta}^{N_{\beta}+1}
n_{\gamma}l_{\gamma}^{N_{\gamma}-1}n_{\delta}l_{\delta}^{N_{\delta}+1})$ \\    \hline
{\bf 31.} & $\beta$   & $\delta$ & $\alpha$ & $\gamma$  & \\
{\bf 32.} & $\delta$  & $\beta$  & $\gamma$ & $\alpha$
& $ (n_{\alpha}l_{\alpha}^{N_{\alpha}}n_{\beta}l_{\beta}^{N_{\beta}}
n_{\gamma}l_{\gamma}^{N_{\gamma}}n_{\delta}l_{\delta}^{N_{\delta}}||$ \\
{\bf 33.} & $\beta$   & $\delta$ & $\gamma$ & $\alpha$  &
$\widehat{G}
(n_il_in_jl_jn_i^{\prime}l_i^{\prime }n_j^{\prime}l_j^{\prime })$ \\
{\bf 34.} & $\delta$  & $\beta$  & $\alpha$ & $\gamma$
& $||n_{\alpha}l_{\alpha}^{N_{\alpha}+1}n_{\beta}l_{\beta}^{N_{\beta}-1}
n_{\gamma}l_{\gamma}^{N_{\gamma}+1}n_{\delta}l_{\delta}^{N_{\delta}-1})$ \\    \hline
{\bf 35.} & $\alpha$  & $\delta$ & $\beta$  & $\gamma$  & \\
{\bf 36.} & $\delta$  & $\alpha$ & $\gamma$ & $\beta$
& $ (n_{\alpha}l_{\alpha}^{N_{\alpha}}n_{\beta}l_{\beta}^{N_{\beta}}
n_{\gamma}l_{\gamma}^{N_{\gamma}}n_{\delta}l_{\delta}^{N_{\delta}}||$ \\
{\bf 37.} & $\alpha$  & $\delta$ & $\gamma$ & $\beta$  &
$\widehat{G}
(n_il_in_jl_jn_i^{\prime}l_i^{\prime }n_j^{\prime}l_j^{\prime })$ \\
{\bf 38.} & $\delta$  & $\alpha$ & $\beta$  & $\gamma$
& $||n_{\alpha}l_{\alpha}^{N_{\alpha}-1}n_{\beta}l_{\beta}^{N_{\beta}+1}
n_{\gamma}l_{\gamma}^{N_{\gamma}+1}n_{\delta}l_{\delta}^{N_{\delta}-1})$ \\    \hline
{\bf 39.} & $\beta$  & $\gamma$ & $\alpha$ & $\delta$  & \\
{\bf 40.} & $\gamma$ & $\beta$  & $\delta$ & $\alpha$
& $ (n_{\alpha}l_{\alpha}^{N_{\alpha}}n_{\beta}l_{\beta}^{N_{\beta}}
n_{\gamma}l_{\gamma}^{N_{\gamma}}n_{\delta}l_{\delta}^{N_{\delta}}||$ \\
{\bf 41.} & $\beta$  & $\gamma$ & $\delta$ & $\alpha$  &
$\widehat{G}
(n_il_in_jl_jn_i^{\prime}l_i^{\prime }n_j^{\prime}l_j^{\prime })$ \\
{\bf 42.} & $\gamma$ & $\beta$  & $\alpha$ & $\delta$
& $||n_{\alpha}l_{\alpha}^{N_{\alpha}+1}n_{\beta}l_{\beta}^{N_{\beta}-1}
n_{\gamma}l_{\gamma}^{N_{\gamma}-1}n_{\delta}l_{\delta}^{N_{\delta}+1})$ \\    \hline
  \end{tabular}
\end{center}

\vspace {0.5in}

%........................................................................

In the work by Gaigalas {\it et al}~\cite{GRF} it is chosen an optimal number
of distributions, which is necessary to obtain the matrix elements of any
two-particle operator, when the bra and ket functions consist of
arbitrary number of shells. This is presented in Table 1.
We point out that for distributions 2-5 and 19-42 the shells' sequence
numbers $\alpha$, $\beta$, $\gamma$, $\delta$ (in bra and ket functions of a
submatrix element) satisfy the condition $\alpha < \beta < \gamma < \delta$,
while for distributions 6-18 no conditions upon
$\alpha$, $\beta$, $\gamma$, $\delta$ are imposed
(This permits to reduce the number of distributions).
For distributions 19-42 this condition is imposed only for obtaining simple
analytical expressions for the recoupling matrices
$R\left( \lambda _i,\lambda _j,\lambda _i^{\prime },\lambda _j^{\prime
},\Lambda ^{bra},\Lambda ^{ket},\Gamma \right)$.
This will be discussed in more detail in the next section.

So, in the way that is described earlier, the Wick's theorem is applied,
assuming that the second quantization operators acting upon shells
$\alpha $, $\beta $, $\gamma$ and $\delta$ belong to different groups.

In the next section we discuss the way to obtain irreducible tensorial form of
these distributions. In addition, the arguments will be given in evidence of
superiority of the obtained tensorial expressions against other
expressions known in the literature.

The methodology presented in this section demonstrates the way to obtain
optimal arrangement of the second quantization operators, for any
two-particle operator. It can be applied without restrictions for
obtaining the optimal tensorial form of two-particle terms of
orthogonal operators and of perturbation theory operators, too.

\section{Graphical methods for two-particle operator}

%\subsection{Generalized graphical method}

The graphical technique of angular momentum is widely used in the atomic
physics: see Yutsis {\it et al}~\cite{YLV},
Jucys and Bandzaitis~\cite{jb}, Brink and Satcher~\cite{brink},
El-Baz~\cite{El-Baz}. It is applied efficiently both in the coordinate
representation (see for example Jucys and Bandzaitis~\cite{jb}),
and in the second quantization formalism (Gaigalas {\it et al}~\cite{ggb}).
The use of it allows one to obtain the analytical expressions for the
recoupling matrices conveniently
(see for example Kaniauskas and Rudzikas \cite{rknj}),
to investigate the tensorial products of operators
(see for example Jucys {\it et al}~\cite{jrb}),
and to seek for the matrix elements of operators
(see for example Huang and Starace~\cite{HS}).
Gaigalas and Merkelis \cite{ggb} have proposed a graphical way to obtain the
values of matrix elements when the operator is a many-particle (one-, two-,
three-, etc.) one and has irreducible tensorial form. For example, when a
two-particle operator is considered, it has the form
(\ref{eq:op-c}) or (\ref{eq:op-e}). The matrix elements in this methodology
are expressed not only in the terms of coefficients of fractional parentage
or reduced coefficients of fractional parentage, but in terms of standard
quantities $U^{k}$ and $V^{k1}$, too. Gaigalas {\it et al}~\cite{GRF}
have proposed to calculate the matrix elements by using the tensorial
expressions for such two-particle operator, that take full advantage of
Racah algebra. In this case the tensorial form of operator depends on the
shells that the operator acts upon (distributions 1-42 from Table~1).
This is the difference of this methodology from other. It is most convenient
to obtain the tensorial expressions for 42 distributions graphically,
using the generalized graphical technique by Gaigalas {\it et al}~\cite{ggb}.
In such a case not only the similarities between different distributions are
easily seen, but also the compact graphical representation of the obtained
expressions is possible. We will stop for details on this in the present
section.

\begin{figure}
\setlength{\unitlength}{1mm}
\begin{picture}(148,180)
\thicklines
%
%Diagram D_{1}
%
\put(10,140){\vector(0,1){10}}
\put(10,150){\vector(0,1){2}}
\put(5,148){\makebox(0,0)[t]{$ n_{i}\lambda_{i}$}}
\put(20,140){\vector(0,1){10}}
\put(20,150){\vector(0,1){2}}
\put(25,148){\makebox(0,0)[t]{$ n_{j}\lambda_{j}$}}
%\multiput(10,90)(1,0){10}{\circle*{0.02}}
\multiput(11,140)(2,0){5}{\oval(2,1)[t]}
\put(15,143){\makebox(0,0){$kk$}}
\put(10,130){\line(0,1){10}}
\put(10,130){\vector(0,1){3}}
\put(5,133){\makebox(0,0){$ n'_{i}\lambda'_{i}$}}
\put(10,128){\vector(0,1){3}}
\put(20,130){\line(0,1){10}}
\put(20,130){\vector(0,1){3}}
\put(25,133){\makebox(0,0){$ n'_{j}\lambda'_{j}$}}
\put(20,128){\vector(0,1){3}}
\put(15,120){\makebox(0,0){$ D_{1}$}}
%
%Diagram D_{2}
%
% linija i
\put(45,140){\line(0,1){12}}
\put(45,140){\circle*{1,7}}
\put(40,148){\makebox(0,0)[t]{$ \lambda_{i}$}}
%linja j
\put(75,140){\line(0,12){12}}
\put(75,152){\line(0,-1){3}}
\put(75,140){\circle*{1,7}}
\put(81,148){\makebox(0,0)[t]{$ \lambda _{j}$}}
% forizontali linija desine
\put(45,140.6){\line(6,0){6}}
\put(45,140.3){\line(6,0){6}}
\put(45,140){\line(20,0){30}}
\put(45,139.7){\line(6,0){6}}
\put(45,139.4){\line(6,0){6}}
\put(52,143){\makebox(0,0){$\kappa_{1} \sigma_{1}$}}
% forizontali linija kaire
\put(69,140.6){\line(6,0){6}}
\put(69,140.3){\line(6,0){6}}
\put(69,139.7){\line(6,0){6}}
\put(69,139.4){\line(6,0){6}}
\put(68,143){\makebox(0,0){$\kappa_{2} \sigma_{2} $}}
%rezultatinis rangas
\put(60,140){\circle*{1,7}}
\put(60,145){\makebox(0,0){${-}$}}
%\put(63,137){\makebox(0,0){${2}$}}
\put(60.6,135){\line(0,1){5}}
\put(60.3,135){\line(0,1){5}}
\put(60,135){\line(0,1){5}}
\put(59.7,135){\line(0,1){5}}
\put(59.4,135){\line(0,1){5}}
\put(60,132){\makebox(0,0){$k k$}}
%linija j (i')
\put(45,130){\line(0,1){10}}
\put(39,134){\makebox(0,0){$ \lambda' _{i}$}}
\put(40,140){\makebox(0,0){${+}$}}
%\put(47,137){\makebox(0,0){${1}$}}
\put(45,130){\line(0,-1){3}}
%linja j'
\put(75,130){\line(0,1){10}}
\put(81,134){\makebox(0,0){$ \lambda'_{j}$}}
\put(80,140){\makebox(0,0){${-}$}}
%\put(73,137){\makebox(0,0){${3}$}}
\put(75,127){\line(0,1){3}}
\put(60,120){\makebox(0,0){$ D_{2}$}}
%
%
%Diagram D_{3}
%
\put(100,136){\vector(0,1){12}}
\put(100,147){\oval(5,5)[b]}
\put(94,144){\makebox(0,0)[t]{$ n_{i}\lambda_{i}$}}
\put(100,120){\makebox(0,0){$ D_{3}$}}
%
%Diagram D_{4}
%
\put(120,126){\line(0,1){10}}
\put(114,130){\makebox(0,0){$ n_{j}\lambda_{j}$}}
\put(120,123){\vector(0,1){3}}
\put(120,125){\oval(5,5)[t]}
\put(120,120){\makebox(0,0){$ D_{4}$}}
%
%
% Diagram D^{5}
%
\put(10,90){\vector(0,1){10}}
\put(10,100){\vector(0,1){2}}
\put(5,98){\makebox(0,0)[t]{$ n_{i}\lambda_{i}$}}
\put(20,90){\vector(0,1){10}}
\put(20,100){\vector(0,1){2}}
\put(25,98){\makebox(0,0)[t]{$ n_{j}\lambda_{j}$}}
\multiput(10,90)(1,0){10}{\circle*{0.02}}
%\multiput(11,90)(2,0){5}{\oval(2,1)[t]}
\put(15,93){\makebox(0,0){$kk$}}
\put(10,80){\line(0,1){10}}
\put(10,80){\vector(0,1){3}}
\put(5,83){\makebox(0,0){$ n'_{i}\lambda'_{i}$}}
\put(10,78){\vector(0,1){3}}
\put(20,80){\line(0,1){10}}
\put(20,80){\vector(0,1){3}}
\put(25,83){\makebox(0,0){$ n'_{j}\lambda'_{j}$}}
\put(20,78){\vector(0,1){3}}
\put(15,70){\makebox(0,0){$ D_{5}$}}
%
%
% Diagram D^{2}
%
%\put(45,90){\vector(0,1){10}}
%\put(45,100){\vector(0,1){2}}
%\put(40,98){\makebox(0,0)[t]{$ n_{i}\lambda_{i}$}}
%\put(55,90){\vector(0,1){10}}
%\put(55,100){\vector(0,1){2}}
%\put(60,98){\makebox(0,0)[t]{$ n_{j}\lambda_{j}$}}
%\multiput(45,90)(1,0){10}{\circle*{0.02}}
%\put(50,93){\makebox(0,0){$kk$}}
%\put(45,80){\line(0,1){10}}
%\put(45,80){\vector(0,1){3}}
%\put(40,83){\makebox(0,0){$ n'_{i}\lambda'_{i}$}}
%\put(45,78){\vector(0,1){3}}
%\put(55,80){\line(0,1){10}}
%\put(55,80){\vector(0,1){3}}
%\put(60,83){\makebox(0,0){$ n'_{j}\lambda'_{j}$}}
%\put(55,78){\vector(0,1){3}}
%\put(50,70){\makebox(0,0){$ D_{2}$}}
%
% Diagram D^{6}
%
% linija i
\put(80,90){\vector(0,1){12}}
\put(80,101){\oval(5,5)[b]}
\put(80,90){\circle*{1,7}}
\put(75,98){\makebox(0,0)[t]{$ n_{i}\lambda_{i}$}}
%linja i' (j)
\put(110,90){\line(0,12){12}}
\put(110,102){\vector(0,-1){3}}
\put(110,101){\oval(5,5)[b]}
\put(110,90){\circle*{1,7}}
\put(115,98){\makebox(0,0)[t]{$ n'_{i}\lambda'_{i}$}}
% forizontali linija desine
\put(80,90.6){\line(6,0){6}}
\put(80,90.3){\line(6,0){6}}
\put(80,90){\line(20,0){30}}
\put(80,89.7){\line(6,0){6}}
\put(80,89.4){\line(6,0){6}}
\put(87,93){\makebox(0,0){$\kappa_{12} \sigma_{12}$}}
% forizontali linija kaire
\put(104,90.6){\line(6,0){6}}
\put(104,90.3){\line(6,0){6}}
\put(104,89.7){\line(6,0){6}}
\put(104,89.4){\line(6,0){6}}
\put(103,93){\makebox(0,0){$\kappa_{12}' \sigma_{12}' $}}
%rezultatinis rangas
\put(95,90){\circle*{1,7}}
\put(95,95){\makebox(0,0){${-}$}}
%\put(98,87){\makebox(0,0){${2}$}}
\put(95.6,85){\line(0,1){5}}
\put(95.3,85){\line(0,1){5}}
\put(95,85){\line(0,1){5}}
\put(94.7,85){\line(0,1){5}}
\put(94.4,85){\line(0,1){5}}
\put(95,82){\makebox(0,0){$k k$}}
%linija j (i')
\put(80,80){\line(0,1){10}}
\put(74,84){\makebox(0,0){$ n_{j}\lambda_{j}$}}
\put(75,90){\makebox(0,0){${+}$}}
%\put(82,87){\makebox(0,0){${1}$}}
\put(80,80){\vector(0,-1){3}}
\put(80,79){\oval(5,5)[t]}
%linja j'
\put(110,80){\line(0,1){10}}
\put(116,84){\makebox(0,0){$ n'_{j}\lambda'_{j}$}}
\put(115,90){\makebox(0,0){${-}$}}
%\put(108,87){\makebox(0,0){${3}$}}
\put(110,77){\vector(0,1){3}}
\put(110,79){\oval(5,5)[t]}
\put(95,70){\makebox(0,0){$ D_{6}$}}
%
% Diagram D^{7}
%
\put(10,40){\vector(0,1){10}}
\put(10,50){\vector(0,1){2}}
\put(5,48){\makebox(0,0)[t]{$ n_{i}\lambda_{i}$}}
%\put(20,40){\vector(0,1){10}}
%\put(20,50){\vector(0,1){2}}
%\put(25,48){\makebox(0,0)[t]{$ n_{j}\lambda_{j}$}}
\multiput(10,40)(1,0){10}{\circle*{0.02}}
\put(15,37){\makebox(0,0){$kk$}}
%\put(10,30){\line(0,1){10}}
%\put(10,30){\vector(0,1){3}}
\put(16,48){\makebox(0,0){$ i'=j$}}
\put(16,45){\vector(-1,0){2}}
\put(20,30){\line(0,1){10}}
\put(15,40){\oval(10,10)[t]}
\put(20,30){\vector(0,1){3}}
\put(26,33){\makebox(0,0){$ n'_{j}\lambda'_{j}$}}
\put(20,28){\vector(0,1){3}}
\put(15,20){\makebox(0,0){$ D_{7}$}}
%
% Diagram D^{8}
%
% linija i
\put(45,40){\vector(0,1){12}}
\put(45,51){\oval(5,5)[b]}
\put(45,40){\circle*{1,7}}
\put(40,48){\makebox(0,0)[t]{$ n_{i}\lambda_{i}$}}
%linja j
\put(75,40){\vector(0,1){12}}
\put(75,51){\oval(5,5)[b]}
\put(75,40){\circle*{1,7}}
\put(80,48){\makebox(0,0)[t]{$ n_{j}\lambda_{j}$}}
% forizontali linija desine
\put(45,40.6){\line(6,0){6}}
\put(45,40.3){\line(6,0){6}}
\put(45,40){\line(20,0){30}}
\put(45,39.7){\line(6,0){6}}
\put(45,39.4){\line(6,0){6}}
\put(52,43){\makebox(0,0){$\kappa_{1} \sigma_{1}$}}
% forizontali linija kaire
\put(69,40.6){\line(6,0){6}}
\put(69,40.3){\line(6,0){6}}
\put(69,39.7){\line(6,0){6}}
\put(69,39.4){\line(6,0){6}}
\put(68,43){\makebox(0,0){$\kappa_{2} \sigma_{2}$}}
%rezultatinis rangas
\put(60,40){\circle*{1,7}}
\put(60,45){\makebox(0,0){${-}$}}
%\put(63,37){\makebox(0,0){${2}$}}
\put(60.6,35){\line(0,1){5}}
\put(60.3,35){\line(0,1){5}}
\put(60,35){\line(0,1){5}}
\put(59.7,35){\line(0,1){5}}
\put(59.4,35){\line(0,1){5}}
\put(60,32){\makebox(0,0){$k k$}}
%linija i'
\put(45,30){\line(0,1){10}}
\put(39,34){\makebox(0,0){$ n'_{i}\lambda'_{i}$}}
\put(40,40){\makebox(0,0){${+}$}}
%\put(42,37){\makebox(0,0){${1}$}}
\put(45,27){\vector(0,1){3}}
\put(45,29){\oval(5,5)[t]}
%linja j'
\put(75,30){\line(0,1){10}}
\put(81,34){\makebox(0,0){$ n'_{j}\lambda'_{j}$}}
\put(80,40){\makebox(0,0){${-}$}}
%\put(73,37){\makebox(0,0){${3}$}}
\put(75,27){\vector(0,1){3}}
\put(75,29){\oval(5,5)[t]}
\put(60,20){\makebox(0,0){$ D_{8}$}}
%
% Diagram D^{9}
%
%rezultatinis rangas
\put(110,42){\circle*{1,7}}
\put(110,37){\makebox(0,0){${+}$}}
\put(110.6,42){\line(0,1){5}}
\put(110.3,42){\line(0,1){5}}
\put(110,42){\line(0,1){5}}
\put(109.7,42){\line(0,1){5}}
\put(109.4,42){\line(0,1){5}}
\put(115,45){\makebox(0,0){$k k$}}
%linija i
\put(100,32){\line(1,1){10}}
\put(100,27){\line(0,1){5}}
\put(94,34){\makebox(0,0){$ n_{i}\lambda_{i}$}}
\put(100,30){\vector(0,-1){3}}
\put(100,29){\oval(5,5)[t]}
%linja j'
\put(120,32){\line(-1,1){10}}
\put(120,27){\line(0,1){5}}
\put(126,34){\makebox(0,0){$ n'_{j}\lambda'_{j}$}}
\put(120,27){\vector(0,1){3}}
\put(120,29){\oval(5,5)[t]}
\put(110,20){\makebox(0,0){$ D_{9}$}}
\end{picture}
\caption{Diagrams for two-particle operators.}
\label{fig:dd-a}
\end{figure}
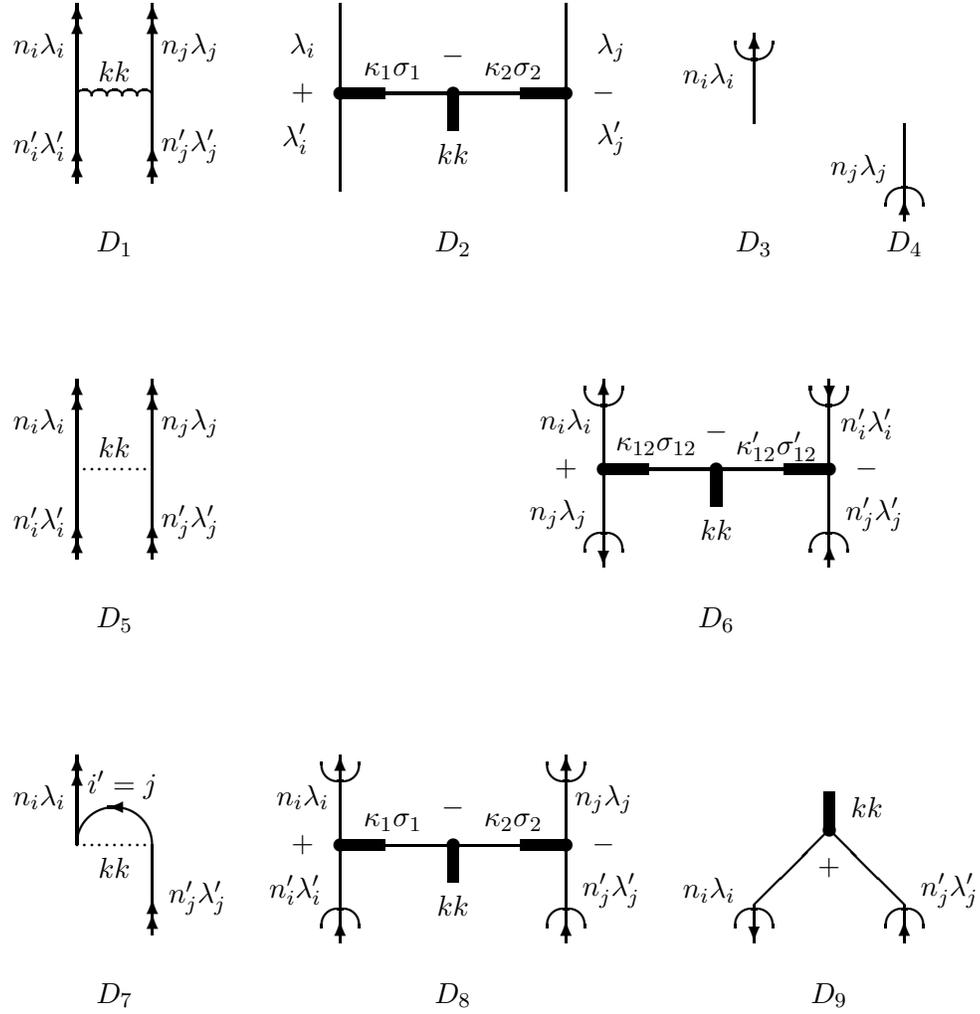

A two-particle operator may be represented graphically by a
Feynman-Goldstone diagram
$D_{1}$ from Figure~\ref{fig:dd-a}
(Lindgren and Morrison~\cite{LM}). As it is shown in the paper
Bolotin {\it et al}~\cite{BLT}, the Feynman-Goldstone
diagrams are topologically
equivalent to the angular momentum graphs. Due to that, an irreducible
tensorial form for every Feynman-Goldstone diagram may be obtained
(see Merkelis {\it et al}~\cite{MGKR}). The graph $D_{2}$ is the angular
momentum graph corresponding to the diagram $D_{1}$. So the two-particle
operator will be written down as follows:

\begin{equation}
\label{eq:pas-a}
\begin{array}[b]{c}
\widehat{G} (ij,i^{\prime }j^{\prime })=
D_{1}= - \frac 12
\displaystyle {\sum_{m_{\lambda _i}m_{\lambda _j}
m_{\lambda ^{\prime }_i}m_{\lambda ^{\prime }_j}}}
\displaystyle {\sum_p}
\left[ \kappa _{1},\kappa _{2},\sigma _{1},\sigma _{2}\right] ^{-1/2}
\times \\ \times \left( n_i\lambda
_in_j\lambda _j||g^{\left( \kappa _1\kappa _2k,\sigma _1\sigma _2k\right)
}||n_i^{\prime }\lambda _i^{\prime }n_j^{\prime }\lambda _j^{\prime }\right)
~D_{2}~~
a^{\left( \lambda _i\right) } _{m_{\lambda _i}}
a^{\left( \lambda _j\right) } _{m_{\lambda _j}}
\stackrel{\sim }{a}^{\left( \lambda _i^{\prime }\right)}
_{m_{\lambda ^{\prime }_i}}
\stackrel{\sim }{a}^{\left( \lambda _j^{\prime }\right)}
_{m_{\lambda ^{\prime }_j}},
\end{array}
\end{equation}

% Figure 1  (begin)
%\ref{fig:dd-a})
%
%\input{a-a}
%

It must be noted that in expression (\ref{eq:pas-a}) the projection
$m_{\lambda _i}$ of $a^{\left( \lambda _i\right) } _{m_{\lambda _i}}$,
as well as that of the momentum line $\lambda _{i}$ in graph $D_{2}$
are the same. This is also to be said about the remaining operators of
second quantization and the three open lines af graph $D_{2}$.

As it has been mentiond in section 2, this operator has two tensorial forms,
(\ref{eq:op-c}) and (\ref{eq:op-d}). These may be represented graphically,
since the creation operator $a^{\left( \lambda _i\right) }$, as well as
operator $\stackrel{\sim }{a}^{\left( \lambda _j^{\prime }\right)}$
respectivelly, are graphically denoted by diagrams $D_{3}$ and $D_{4}$.

The first form (\ref{eq:op-c}) of two-particle operator $G_I(ij,i^{\prime
}j^{\prime })$ is represented as:

\begin{equation}
\label{eq:pas-b}
\begin{array}[b]{c}
G_I(ij,i^{\prime }j^{\prime })=D_1=-\frac 12
\displaystyle {\sum_{\kappa _{12}\kappa _{12}^{^{\prime }}\sigma _{12}\sigma
_{12}^{^{\prime }}}}\displaystyle {\sum_p}\left( -1\right) ^{k-p}\left[
\kappa _1,\kappa _2,\sigma _1,\sigma _2\right] ^{1/2}\times \\ \times \left(
n_i\lambda _in_j\lambda _j||g^{\left( \kappa _1\kappa _2k,\sigma _1\sigma
_2k\right) }||n_i^{\prime }\lambda _i^{\prime }n_j^{\prime }\lambda
_j^{\prime }\right) \left\{
\begin{array}{ccc}
l_i^{\prime } & l_j^{\prime } & \kappa _{12}^{\prime } \\
\kappa _1 & \kappa _2 & k \\
l_i & l_j & \kappa _{12}
\end{array}
\right\} \times \\
\times \left\{
\begin{array}{ccc}
s & s & \sigma _{12}^{\prime } \\
\sigma _1 & \sigma _2 & k \\
s & s & \sigma _{12}
\end{array}
\right\} D_6
\end{array}
\end{equation}
whereas the second (\ref{eq:op-d}):

\begin{equation}
\label{eq:pas-c}
\begin{array}[b]{c}
G_{II}(ij,i^{\prime }j^{\prime })=D_5+D_7= \\
=\frac 12
\displaystyle {\sum_p}\left( -1\right) ^{k-p}\left( n_i\lambda _in_j\lambda
_j||g^{\left( \kappa _1\kappa _2k,\sigma _1\sigma _2k\right) }||n_i^{\prime
}\lambda _i^{\prime }n_j^{\prime }\lambda _j^{\prime }\right) \times \\
\times \{\left[ \kappa _1,\kappa _2,\sigma _1,\sigma _2\right]
^{-1/2}D_8-\left( -1\right) ^{l_i+l_j^{\prime }}\left\{
\begin{array}{ccc}
\kappa _1 & \kappa _2 & k \\
l_j^{\prime } & l_i & l_j
\end{array}
\right\} \times \\
\times \left\{
\begin{array}{ccc}
\sigma _1 & \sigma _2 & k \\
s & s & s
\end{array}
\right\} \delta \left( n_jl_j,n_i^{\prime }l_i^{\prime }\right) D_9\}.
\end{array}
\end{equation}

We emphasize here that the winding line of interaction in the
Feynman-Goldstone diagram corresponds to the operators of second
quantization in the normal order (Figure~\ref{fig:dd-a}, $D_1$). Whereas the
dotted interaction line indicates that the second quantization operators are
ordered in pairs of creation-annihilation operators. In the latter case first
comes
the pair on the left side of a Feynman-Goldstone diagram
(Figure~\ref{fig:dd-a}, $D_5$).
Such a notation of two kinds for an interaction line is meaningful only for
two-particle (or more) operators, since for any one-particle operator both
the winding and dotted lines correspond to the same order of creation and
annihilation operators.

From expressions (\ref{eq:pas-b}), (\ref{eq:pas-c}) we see that the
two-particle operator in the first form is represented by one
Feynman-Goldstone diagram $D_1$, whereas in the second - by two diagrams $D_5$
and $D_7$. The diagrams, corresponding to tensorial product, have the
following algebraic expressions:

\begin{equation}
\label{eq:pas-d}D_6=\left[ \left[ a^{\left( \lambda _i\right) }\times
a^{\left( \lambda _j\right) }\right] ^{\left( \kappa _{12}\sigma
_{12}\right) }\times \left[ \stackrel{\sim }{a}^{\left( \lambda _i^{\prime
}\right) }\times \stackrel{\sim }{a}^{\left( \lambda _j^{\prime }\right)
}\right] ^{\left( \kappa _{12}^{\prime }\sigma _{12}^{\prime }\right)
}\right] _{p-p}^{\left( kk\right) },
\end{equation}

\begin{equation}
\label{eq:pas-e}D_8=\left[ \left[ a^{\left( \lambda _i\right) }\times
\stackrel{\sim }{a}^{\left( \lambda _i^{\prime }\right) }\right] ^{\left(
\kappa _1\sigma _1\right) }\times \left[ a^{\left( \lambda _j\right) }\times
\stackrel{\sim }{a}^{\left( \lambda _j^{\prime }\right) }\right] ^{\left(
\kappa _2\sigma _2\right) }\right] _{p-p}^{\left( kk\right) }
\end{equation}

\begin{equation}
\label{eq:pas-f}D_9=\left[ a^{\left( \lambda _i\right) }\times \stackrel{\sim
}{a}^{\left( \lambda _j^{\prime }\right) }\right] _{p-p}^{\left( kk\right) }
\end{equation}

The positions of the second quantization operators in the diagrams
$D_6$, $D_8$ and $D_9$ define their order in tensorial
products: the first place in tensorial product occupies the upper right
second quantization operator, the second - lower right, after them the upper
left and lower left operators follow. The angular momenta diagram defines
their coupling scheme into tensorial product. For more detail see
Gaigalas and Rudzikas~\cite{GR}.

As it has been mentioned earlier, these two forms do not always take full
advantage of the Racah algebra (see Gaigalas and Rudzikas~\cite{GR}).
The expression (\ref{eq:op-i}) has no such shortcomings. Now we will
demonstrate the way to obtain graphically a tensorial expressions for
particular distributions 1-42 from Table~1.

% Figure 2  (begin)
%
%\input{a-a}
%
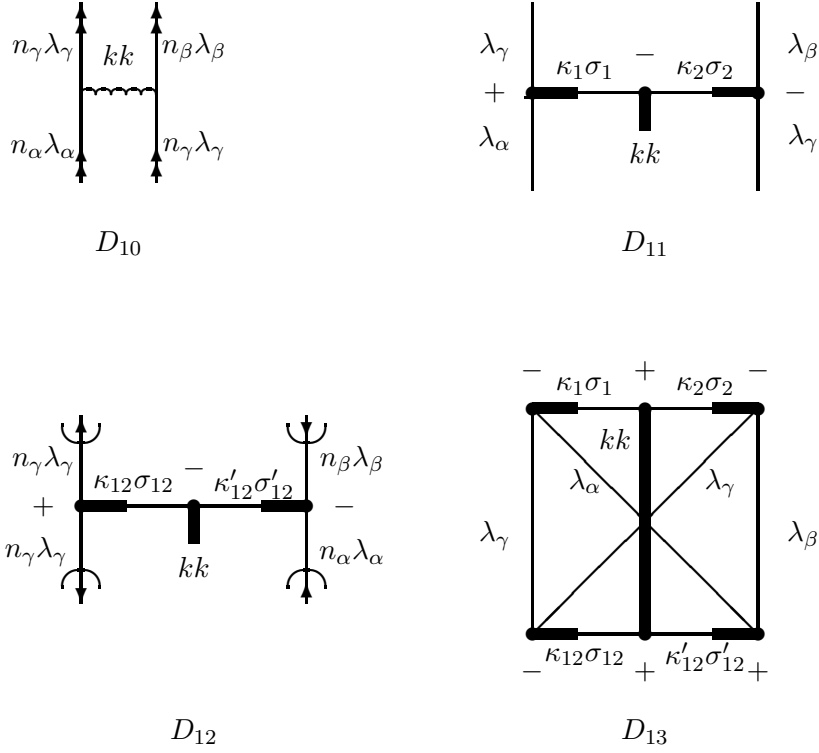
\begin{figure}
\setlength{\unitlength}{1mm}
\begin{picture}(148,120)
\thicklines
%
%Diagram D_{10}
%
\put(10,105){\vector(0,1){10}}
\put(10,115){\vector(0,1){2}}
\put(5,113){\makebox(0,0)[t]{$ n_{\gamma}\lambda_{\gamma}$}}
\put(20,105){\vector(0,1){10}}
\put(20,115){\vector(0,1){2}}
\put(25,113){\makebox(0,0)[t]{$ n_{\beta}\lambda_{\beta}$}}
%\multiput(10,90)(1,0){10}{\circle*{0.02}}
\multiput(11,105)(2,0){5}{\oval(2,1)[t]}
\put(15,110){\makebox(0,0){$kk$}}
\put(10,95){\line(0,1){10}}
\put(10,95){\vector(0,1){3}}
\put(5,98){\makebox(0,0){$ n_{\alpha}\lambda_{\alpha}$}}
\put(10,93){\vector(0,1){3}}
\put(20,95){\line(0,1){10}}
\put(20,95){\vector(0,1){3}}
\put(25,98){\makebox(0,0){$ n_{\gamma}\lambda_{\gamma}$}}
\put(20,93){\vector(0,1){3}}
\put(15,85){\makebox(0,0){$ D_{10}$}}
%
%Diagram D_{11}
%
% linija i
\put(70,105){\line(0,1){12}}
\put(70,105){\circle*{1,7}}
\put(65,113){\makebox(0,0)[t]{$ \lambda_{\gamma}$}}
%linja j
\put(100,105){\line(0,12){12}}
\put(100,117){\line(0,-1){3}}
\put(100,105){\circle*{1,7}}
\put(106,113){\makebox(0,0)[t]{$ \lambda _{\beta}$}}
% forizontali linija desine
\put(70,105.6){\line(6,0){6}}
\put(70,105.3){\line(6,0){6}}
\put(70,105){\line(20,0){30}}
\put(70,104.7){\line(6,0){6}}
\put(70,104.4){\line(6,0){6}}
\put(77,108){\makebox(0,0){$\kappa_{1} \sigma_{1}$}}
% forizontali linija kaire
\put(94,105.6){\line(6,0){6}}
\put(94,105.3){\line(6,0){6}}
\put(94,104.7){\line(6,0){6}}
\put(69,104.4){\line(6,0){6}}
\put(93,108){\makebox(0,0){$\kappa_{2} \sigma_{2} $}}
%rezultatinis rangas
\put(85,105){\circle*{1,7}}
\put(85,110){\makebox(0,0){${-}$}}
%\put(88,102){\makebox(0,0){${2}$}}
\put(85.6,100){\line(0,1){5}}
\put(85.3,100){\line(0,1){5}}
\put(85,100){\line(0,1){5}}
\put(84.7,100){\line(0,1){5}}
\put(84.4,100){\line(0,1){5}}
\put(85,97){\makebox(0,0){$k k$}}
%linija j (i')
\put(70,95){\line(0,1){10}}
\put(65,99){\makebox(0,0){$ \lambda _{\alpha}$}}
\put(65,105){\makebox(0,0){${+}$}}
%\put(72,102){\makebox(0,0){${1}$}}
\put(70,95){\line(0,-1){3}}
%linja j'
\put(100,95){\line(0,1){10}}
\put(106,99){\makebox(0,0){$ \lambda _{\gamma}$}}
\put(105,105){\makebox(0,0){${-}$}}
%\put(73,102){\makebox(0,0){${3}$}}
\put(100,92){\line(0,1){3}}
\put(85,85){\makebox(0,0){$ D_{11}$}}
%
% Diagram D^{12}
%
% linija i
\put(10,50){\vector(0,1){12}}
\put(10,61){\oval(5,5)[b]}
\put(10,50){\circle*{1,7}}
\put(5,58){\makebox(0,0)[t]{$ n_{\gamma}\lambda _{\gamma}$}}
%linja i' (j)
\put(40,50){\line(0,12){12}}
\put(40,62){\vector(0,-1){3}}
\put(40,61){\oval(5,5)[b]}
\put(40,50){\circle*{1,7}}
\put(46,58){\makebox(0,0)[t]{$ n_{\beta}\lambda _{\beta}$}}
% horizontali linija desine
\put(10,50.6){\line(6,0){6}}
\put(10,50.3){\line(6,0){6}}
\put(10,50){\line(20,0){30}}
\put(10,49.7){\line(6,0){6}}
\put(10,49.4){\line(6,0){6}}
\put(17,53){\makebox(0,0){$\kappa_{12} \sigma_{12}$}}
% horizontali linija kaire
\put(34,50.6){\line(6,0){6}}
\put(34,50.3){\line(6,0){6}}
\put(34,49.7){\line(6,0){6}}
\put(34,49.4){\line(6,0){6}}
\put(33,53){\makebox(0,0){$\kappa_{12}' \sigma_{12}' $}}
%rezultatinis rangas
\put(25,50){\circle*{1,7}}
\put(25,55){\makebox(0,0){${-}$}}
%\put(28,47){\makebox(0,0){${2}$}}
\put(25.6,45){\line(0,1){5}}
\put(25.3,45){\line(0,1){5}}
\put(25,45){\line(0,1){5}}
\put(24.7,45){\line(0,1){5}}
\put(24.4,45){\line(0,1){5}}
\put(25,42){\makebox(0,0){$k k$}}
%linija j (i')
\put(10,40){\line(0,1){10}}
\put(4,44){\makebox(0,0){$ n_{\gamma}\lambda _{\gamma}$}}
\put(5,50){\makebox(0,0){${+}$}}
%\put(12,47){\makebox(0,0){${1}$}}
\put(10,40){\vector(0,-1){3}}
\put(10,39){\oval(5,5)[t]}
%linja j'
\put(40,40){\line(0,1){10}}
\put(46,44){\makebox(0,0){$ n_{\alpha}\lambda _{\alpha}$}}
\put(45,50){\makebox(0,0){${-}$}}
%\put(38,47){\makebox(0,0){${3}$}}
\put(40,37){\vector(0,1){3}}
\put(40,39){\oval(5,5)[t]}
\put(25,20){\makebox(0,0){$ D_{12}$}}
%
%
% Diagram D^{13}
%
\put(70,63){\circle*{1,7}}
\put(100,63){\circle*{1,7}}
% istrizine linia 1
\put(70,63){\line(1,-1){30}}
% istrizine linia 1
\put(100,63){\line(-1,-1){30}}
% horizontali linija desine
\put(70,63.6){\line(6,0){6}}
\put(70,63.3){\line(6,0){6}}
\put(70,63){\line(20,0){30}}
\put(70,62.7){\line(6,0){6}}
\put(70,62.4){\line(6,0){6}}
\put(77,66){\makebox(0,0){$\kappa_{1} \sigma_{1}$}}
% horizontali linija kaire
\put(94,63.6){\line(6,0){6}}
\put(94,63.3){\line(6,0){6}}
\put(94,62.7){\line(6,0){6}}
\put(94,62.4){\line(6,0){6}}
\put(93,66){\makebox(0,0){$\kappa_{2} \sigma_{2} $}}
%rezultatinis rangas
\put(85,63){\circle*{1,7}}
\put(85,68){\makebox(0,0){${+}$}}
%\put(88,60){\makebox(0,0){${2}$}}
\put(85.6,63){\line(0,-1){30}}
\put(85.3,63){\line(0,-1){30}}
\put(85,63){\line(0,-1){30}}
\put(84.7,63){\line(0,-1){30}}
\put(84.4,63){\line(0,-1){30}}
\put(81,59){\makebox(0,0){$k k$}}
%vertikali linija 1
\put(77,55){\makebox(0,0)[t]{$ \lambda_{\alpha}$}}
\put(65,48){\makebox(0,0)[t]{$ \lambda_{\gamma}$}}
\put(70,68){\makebox(0,0){${-}$}}
\put(70,63){\line(0,-1){30}}
%vertikali linija 2
\put(95,55){\makebox(0,0)[t]{$ \lambda_{\gamma}$}}
\put(106,48){\makebox(0,0)[t]{$ \lambda _{\beta}$}}
\put(100,68){\makebox(0,0){${-}$}}
\put(100,63){\line(0,-1){30}}
%----------------------------------
\put(85,33){\circle*{1,7}}
\put(85,28){\makebox(0,0){${+}$}}
% horizontali linija desine
\put(70,28){\makebox(0,0){${-}$}}
\put(70,33){\circle*{1,7}}
\put(70,33.6){\line(6,0){6}}
\put(70,33.3){\line(6,0){6}}
\put(70,33){\line(20,0){30}}
\put(70,32.7){\line(6,0){6}}
\put(70,32.4){\line(6,0){6}}
\put(77,30){\makebox(0,0){$\kappa_{12} \sigma_{12}$}}
% horizontali linija kaire
\put(100,28){\makebox(0,0){${+}$}}
\put(100,33){\circle*{1,7}}
\put(94,33.6){\line(6,0){6}}
\put(94,33.3){\line(6,0){6}}
\put(94,32.7){\line(6,0){6}}
\put(94,32.4){\line(6,0){6}}
\put(93,30){\makebox(0,0){$\kappa_{12}' \sigma_{12}' $}}
\put(85,20){\makebox(0,0){$ D_{13}$}}
\end{picture}
\caption{Diagrams for distribution $\gamma$, $\beta$, $\alpha$, $\gamma$.}
\label{fig:dd-b}
\end{figure}

We take the distribution $\gamma$, $\beta$, $\alpha$, $\gamma$ (13 form Table 1)
as an example for investigation. Then the Feynman-Goldstone diagram of operator
$\widehat{G}(\gamma \beta , \alpha \gamma )$ is $D_{10}$, the angular
momentum graph is $D_{11}$, and the second quantization operators are in the
following order:
\begin{equation}
\label{eq:pas-g}
\begin{array}{c}
a_{\gamma}a_{\beta}a^{\dagger}_{\alpha}a^{\dagger}_{\gamma}.
\end{array}
\end{equation}
Applying the Wick's theorem as described in Section 3, and assuming that
the operators acting upon shell $\gamma$ belong to the first group, the
once acting upon $\beta$ belong to the second, and the ones acting upon
$\alpha$ belong to the third group, we obtain the following order of operators:
\begin{equation}
\label{eq:pas-h}
\begin{array}{c}
a_{\gamma}a^{\dagger}_{\gamma}a_{\beta}a^{\dagger}_{\alpha}.
\end{array}
\end{equation}
Using the generalized graphical technique of angular momentum by
Gaigalas {\it et al}~\cite{gga}, we couple the operators of second quantization
into tensorial product $D_{12}$ (see Figure~\ref{fig:dd-c}):

\begin{equation}
\label{eq:pas-i}D_{12}=\left[ \left[ a^{\left( \lambda _{\gamma}\right) }\times
a^{\left( \lambda _{\gamma}\right) }\right] ^{\left( \kappa _{12}\sigma
_{12}\right) }\times \left[ \stackrel{\sim }{a}^{\left( \lambda _{\beta}
\right) }\times \stackrel{\sim }{a}^{\left( \lambda _{\alpha}\right)
}\right] ^{\left( \kappa _{12}^{\prime }\sigma _{12}^{\prime }\right)
}\right] _{p-p}^{\left( kk\right) }.
\end{equation}
In the course of obtaining $D_{12}$ graphically, a recoupling matrix $D_{13}$
appears, whose analytical expression is readily obtained by using the
graphical technique of Jucys and Bandzaitis~\cite{jb}. All the needed
expressions are obtained in the same way.

%
% Figure 3  (begin)
%
%input{c-a}
%
\begin{figure}
\setlength{\unitlength}{1mm}
\begin{picture}(148,105)
\thicklines
% Diagram D_{14}
\put(20,90){\vector(0,1){10}}
\put(20,100){\vector(0,1){2}}
\put(15,98){\makebox(0,0)[t]{$ n_{3}\lambda_{3}$}}
\put(30,90){\vector(0,1){10}}
\put(30,100){\vector(0,1){2}}
\put(35,98){\makebox(0,0)[t]{$ n_{1}\lambda_{1}$}}
\multiput(20,90)(1,0){10}{\circle*{0.02}}
\put(25,93){\makebox(0,0){$kk$}}
\put(20,80){\line(0,1){10}}
\put(20,80){\vector(0,1){3}}
\put(15,83){\makebox(0,0){$ n_{4}\lambda_{4}$}}
\put(20,78){\vector(0,1){3}}
\put(30,80){\line(0,1){10}}
\put(30,80){\vector(0,1){3}}
\put(35,83){\makebox(0,0){$ n_{2}\lambda_{2}$}}
\put(30,78){\vector(0,1){3}}
\put(25,70){\makebox(0,0){$ D_{14}$}}
%
% Diagram D_{15}
% line i
\put(80,90){\vector(0,1){12}}
\put(80,101){\oval(5,5)[b]}
\put(80,90){\circle*{1,7}}
\put(74,98){\makebox(0,0)[t]{$ n_{3}\lambda_{3}$}}
%line j
\put(110,90){\vector(0,1){12}}
\put(110,101){\oval(5,5)[b]}
\put(110,90){\circle*{1,7}}
\put(115,98){\makebox(0,0)[t]{$ n_{1}\lambda_{1}$}}
% forizontali linija desine
\put(80,90.6){\line(6,0){6}}
\put(80,90.3){\line(6,0){6}}
\put(80,90){\line(20,0){30}}
\put(80,89.7){\line(6,0){6}}
\put(80,89.4){\line(6,0){6}}
\put(87,93){\makebox(0,0){$\kappa_{1} \sigma_{1}$}}
% forizontali linija kaire
\put(104,90.6){\line(6,0){6}}
\put(104,90.3){\line(6,0){6}}
\put(104,89.7){\line(6,0){6}}
\put(104,89.4){\line(6,0){6}}
\put(103,93){\makebox(0,0){$\kappa_{2} \sigma_{2}$}}
%rezultatinis rangas
\put(95,90){\circle*{1,7}}
\put(95,95){\makebox(0,0){${-}$}}
\put(95.6,85){\line(0,1){5}}
\put(95.3,85){\line(0,1){5}}
\put(95,85){\line(0,1){5}}
\put(94.7,85){\line(0,1){5}}
\put(94.4,85){\line(0,1){5}}
\put(95,82){\makebox(0,0){$k k$}}
%line i'
\put(80,80){\line(0,1){10}}
\put(74,84){\makebox(0,0){$ n_{4}\lambda_{4}$}}
\put(75,90){\makebox(0,0){${+}$}}
\put(80,77){\vector(0,1){3}}
\put(80,79){\oval(5,5)[t]}
%line j'
\put(110,80){\line(0,1){10}}
\put(116,84){\makebox(0,0){$ n_{2}\lambda_{2}$}}
\put(115,90){\makebox(0,0){${-}$}}
\put(110,77){\vector(0,1){3}}
\put(110,79){\oval(5,5)[t]}
\put(95,70){\makebox(0,0){$ D_{15}$}}
%
% Diagram D_{16}
% line i
\put(10,40){\vector(0,1){12}}
\put(10,51){\oval(5,5)[b]}
\put(10,40){\circle*{1,7}}
\put(4,48){\makebox(0,0)[t]{$ n_{1}\lambda_{1}$}}
%line j
\put(40,40){\vector(0,1){12}}
\put(40,51){\oval(5,5)[b]}
\put(40,40){\circle*{1,7}}
\put(45,48){\makebox(0,0)[t]{$ n_{3}\lambda_{3}$}}
% forizontali linija desine
\put(10,40.6){\line(6,0){6}}
\put(10,40.3){\line(6,0){6}}
\put(10,40){\line(20,0){30}}
\put(10,39.7){\line(6,0){6}}
\put(10,39.4){\line(6,0){6}}
\put(17,43){\makebox(0,0){$\kappa_{1} \sigma_{1}$}}
% forizontali linija kaire
\put(34,40.6){\line(6,0){6}}
\put(34,40.3){\line(6,0){6}}
\put(34,39.7){\line(6,0){6}}
\put(34,39.4){\line(6,0){6}}
\put(33,43){\makebox(0,0){$\kappa_{2} \sigma_{2}$}}
%rezultatinis rangas
\put(25,40){\circle*{1,7}}
\put(25,45){\makebox(0,0){${+}$}}
\put(28,37){\makebox(0,0){${a}$}}
\put(25.6,35){\line(0,1){5}}
\put(25.3,35){\line(0,1){5}}
\put(25,35){\line(0,1){5}}
\put(24.7,35){\line(0,1){5}}
\put(24.4,35){\line(0,1){5}}
\put(25,32){\makebox(0,0){$k k$}}
%line i'
\put(10,30){\line(0,1){10}}
\put(4,34){\makebox(0,0){$ n_{2}\lambda_{2}$}}
\put(5,40){\makebox(0,0){${+}$}}
\put(10,27){\vector(0,1){3}}
\put(10,29){\oval(5,5)[t]}
%line j'
\put(40,30){\line(0,1){10}}
\put(46,34){\makebox(0,0){$ n_{4}\lambda_{4}$}}
\put(45,40){\makebox(0,0){${-}$}}
\put(40,27){\vector(0,1){3}}
\put(40,29){\oval(5,5)[t]}
\put(25,20){\makebox(0,0){$ D_{16}$}}
%
% Diagram D_{17}
% line i
\put(80,40){\vector(0,1){12}}
\put(80,51){\oval(5,5)[b]}
\put(80,40){\circle*{1,7}}
\put(74,48){\makebox(0,0)[t]{$ n_{1}\lambda_{1}$}}
%linja j
\put(110,40){\vector(0,1){12}}
\put(110,51){\oval(5,5)[b]}
\put(110,40){\circle*{1,7}}
\put(115,48){\makebox(0,0)[t]{$ n_{3}\lambda_{3}$}}
% forizontali linija desine
\put(80,40.6){\line(6,0){6}}
\put(80,40.3){\line(6,0){6}}
\put(80,40){\line(20,0){30}}
\put(80,39.7){\line(6,0){6}}
\put(80,39.4){\line(6,0){6}}
\put(87,43){\makebox(0,0){$\kappa_{1} \sigma_{1}$}}
% forizontali linija kaire
\put(104,40.6){\line(6,0){6}}
\put(104,40.3){\line(6,0){6}}
\put(104,39.7){\line(6,0){6}}
\put(104,39.4){\line(6,0){6}}
\put(103,43){\makebox(0,0){$\kappa_{2} \sigma_{2}$}}
%rezultatinis rangas
\put(95,40){\circle*{1,7}}
\put(95,45){\makebox(0,0){${-}$}}
\put(98,37){\makebox(0,0){${a}$}}
\put(95.6,35){\line(0,1){5}}
\put(95.3,35){\line(0,1){5}}
\put(95,35){\line(0,1){5}}
\put(94.7,35){\line(0,1){5}}
\put(94.4,35){\line(0,1){5}}
\put(95,32){\makebox(0,0){$k k$}}
%line i'
\put(80,30){\line(0,1){10}}
\put(74,34){\makebox(0,0){$ n_{2}\lambda_{2}$}}
\put(75,40){\makebox(0,0){${+}$}}
\put(80,27){\vector(0,1){3}}
\put(80,29){\oval(5,5)[t]}
%line j'
\put(110,30){\line(0,1){10}}
\put(116,34){\makebox(0,0){$ n_{4}\lambda_{4}$}}
\put(115,40){\makebox(0,0){${-}$}}
\put(110,27){\vector(0,1){3}}
\put(110,29){\oval(5,5)[t]}
\put(95,20){\makebox(0,0){$ D_{17}$}}
\end{picture}
\caption{Diagrams for graphical transformations.}
\label{fig:dd-c}
\end{figure}
%
% Figure 3  (end)
%

In obtaining these expressions, as well as in representing them graphically,
it is very convenient to use the rule of changing the sign of a node,
existing in the graphical technique of angular momentum
(see Jucys and Bandzaitis~\cite{jb}). Therefore now we will treat an example
of using such a rule.

Suppose, we have the following correspondence between diagrams
(Figure~\ref{fig:dd-c}):

\begin{equation}
\label{eq:grf-a}D_{14}\longrightarrow D_{15},
\end{equation}
in which the second quantization operators are in the order $a^{(\lambda
_3)} $ $\tilde a^{(\lambda _4)}$ $a^{(\lambda _1)}$ $\tilde a^{(\lambda _2)}$%
. Our goal is to obtain the diagram corresponding to the order $a^{(\lambda
_1)}$ $\tilde a^{(\lambda _2)}$ $a^{(\lambda _3)}$ $\tilde a^{(\lambda _4)}$%
. Bearing in mind that the second quantization operators anticommute with
each other and they all act on different electronic shells and we are not
changing the order of their coupling into tensorial product, we arrive at

\begin{equation}
\label{eq:grf-b}D_{14}\longrightarrow (-1)^4D_{16}=D_{16}.
\end{equation}

Let us also discuss another situation: we have defined the disposition of
the operators and we want to couple them into certain tensorial product.
Suppose that we want to represent graphically the following tensorial
product:

\begin{equation}
\label{eq:grf-c}\left[ \left[ a^{(\lambda _1)}\times \tilde a^{(\lambda
_2)}\right] ^{\left( \kappa _1\sigma _1\right) }\times \left[ a^{(\lambda
_3)}\times \tilde a^{(\lambda _4)}\right] ^{\left( \kappa _2\sigma _2\right)
}\right] ^{\left( \kappa \sigma \right) }.
\end{equation}

For this purpose we have to rearrange the generalized Clebsch-Gordan
coefficient, which is defining the scheme of coupling of the operators into
the tensorial product. It is easy to notice that this coefficient will
properly define the tensorial product, if we change the sign of the vertex ''%
$a$'' in diagram $D_{16}$:

\begin{equation}
\label{eq:grf-d}D_{14}\longrightarrow (-1)^{\kappa _1+\kappa _2-\kappa +\sigma
_1+\sigma _2-\sigma }D_{17}.
\end{equation}

The procedures described are fairly simple, however, they are sufficient for
the majority of cases.
The more complete description of this generalized
graphical approach may be found in Gaigalas {\it et al}~\cite{gga},
Gaigalas~\cite{gg}, Gaigalas and Merkelis~\cite{ggb}.

All the analytical expressions for distributions 1 - 42 from Table 1
are presented in the paper Gaigalas {\it et al}~\cite{GRF}. They are written
down using the generalized graphical methodology of angular momentum, and
the vortex sign change rule, which was discussed in this section. As a
consequence of that, the analytical expressions for 42 terms may be written
down via 6 different expressions. This, undoubtedly, facilitates a lot the
implementation of methodology proposed in
Gaigalas {\it et al}~\cite{GRF}.

\section{ Matrix Elements Between Complex Configurations}

In this section we will discuss several ways to obtain matrix elements of
a two-particle operator.
As it was mentioned earlier, up to now the Fano
calculation scheme~\cite{Fano} is
the most popular one. Its general expression when a two-particle operator
acts upon different shells is presented in (\ref{eq:ma-c}).

The general expression for a matrix element in other cases is similar.
For example, when the operator acts only upon one shell, we have

\begin{equation}
\label{eq:ma-d}
\begin{array}[b]{c}
(\psi _u^{bra}\left( LS\right) ||
\widehat{G}^{\left( \kappa _1\kappa _2k,\sigma_1\sigma _2k\right) }
||\psi _u^{ket}\left( L^{\prime }S^{\prime}\right) ) \sim \\
\sim \displaystyle {\sum_{n_{i} \lambda _{i} }}
\left( -1\right) ^\Delta N_{i} \left( N_{i}-1 \right) \times \\
\times
\displaystyle {\sum_{ \{ T \} }}
\left( l_{i}^{N_{i}}\;\alpha_{i} L_{i}S_{i}||l_{i}^{N_{i}-1}\;\left(
\alpha_{i}^{\prime \prime} L^{\prime \prime}_{i} S^{\prime \prime}_{i}
\right), l_{i} \right)
\left( l_{i}^{N_{i}-1}\;\alpha_{i}^{\prime \prime}
L_{i}^{\prime \prime}S_{i}^{\prime \prime}
||l_{j}^{N_{i}-2}\;\left(
\alpha_{i}^{\prime \prime \prime} L_{i}^{\prime \prime \prime}
S^{\prime \prime \prime}_{i} \right), l_{i} \right) \times  \\
\times \left( l_{i}^{N_{i}^{\prime }}\;
\alpha_{i}^{\prime \prime \prime}
L_{i}^{\prime \prime \prime}S_{i}^{\prime \prime \prime}
||l_{i}^{N_{i}^{\prime }-1}\;\left(
\alpha_{i}^{\prime V} L^{\prime V}_{i}
S^{\prime V}_{i} \right), l_{i} \right)
\left( l_{i}^{N_{i}^{\prime }-1}\;\alpha_{i}^{\prime V}
L_{i}^{\prime V}S_{i}^{\prime V}||l_{i}^{N_{i}^{\prime }-2}\;\left(
\alpha_{i}^{\prime } L^{\prime}_{i}
S^{\prime }_{i} \right), l_{i} \right) \times  \\
\times
R_{d} \left( \lambda _i,\lambda _i,\lambda _i,\lambda _i,
\kappa_1, \kappa_2, \kappa, \sigma_1, \sigma_2, \sigma, \Lambda ^{bra},
\Lambda ^{ket} \right)  \times \\
\times \left( n_i\lambda _in_i\lambda _i||
g^{\left( \kappa _1\kappa _2k,\sigma_1\sigma _2k\right) }
||n_i\lambda _in_i\lambda_i\right).
\end{array}
\end{equation}

As we see here, in contrast to (\ref{eq:ma-c}),
the summation is performed only over one array $n$, $\lambda$
of quantum numbers, because the operator acts only upon one shell. But here
a summation over $\{T\}$ occurs, though, which indicates the summation
over arrays of intermediate terms
$\alpha_{i}^{\prime \prime} L^{\prime \prime}_{i} S^{\prime \prime}_{i}$,
$\alpha_{i}^{\prime \prime \prime} L^{\prime \prime \prime}_{i}
S^{\prime \prime \prime}_{i}$,
$\alpha_{i}^{\prime V} L^{\prime V}_{i} S^{\prime V}_{i}$.

Remembering the relationship between a coefficient of fractional parentage
and a reduced matrix element of a second quantization operator
(see \v Spakauskas~{\it et al} \cite{SKRb}, Rudzikas and
Kaniauskas~\cite{ra}):
\begin{equation}
\label{eq:ma-e}
\begin{array}[b]{c}
\left( l^{N}\;\alpha LS||a^{(ls)}|| l^{N-1}\;
\alpha^{\prime } L^{\prime } S^{\prime } \right) = \\
= (-1)^{N+(N+1)\varphi (N)} \sqrt{N[L,S]}
\left( l^{N}\;\alpha LS||l^{N-1}\;\left(
\alpha^{\prime } L^{\prime } S^{\prime } \right), l \right)
\end{array}
\end{equation}
we see that the Racah algebra in expressions (\ref{eq:ma-c}), (\ref{eq:ma-d})
is used only on the level of coefficients of fractional parentage. In separate
cases, e.g. when the two-particle operator acts upon one or two shells, it
is possible to use expressions  which exploit the Racah algebra at a higher
level, i.e. to take more advantage of the tensor algebra elements
(see Judd~\cite{Judd}, Jucys and Savukynas~\cite{js}).
For example, let us investigate the case when a matrix element is calculated
for bra and ket functions having one shell only. The tensorial forms
(\ref{eq:op-c}) and (\ref{eq:op-e}) are of value here. Taking the second one
of these, we have
\begin{equation}
\label{eq:ma-f}
\begin{array}[b]{c}
(nl^{N}\alpha LS|\widehat{G}_{II}|nl^{N}\alpha ^{\prime}L^{\prime}S^{\prime})= \\
=\frac 12 \left( n\lambda n\lambda
||g^{\left( \kappa _1\kappa _2k,\sigma _1\sigma _2k\right) }||n\lambda
n\lambda \right)
\left\{ \left[ \kappa _1,\kappa _2,\sigma_1,\sigma _2\right] ^{-1/2} \right.
\times \\
\times (nl^{N}\alpha LS|
\left[ \left[ a^{\left( \lambda \right) }\times
\stackrel{\sim }{a}^{\left( \lambda \right) }\right] ^{\left( \kappa
_1\sigma _1\right) }\times \left[ a^{\left( \lambda \right) }\times
\stackrel{\sim }{a}^{\left( \lambda \right) }\right] ^{\left( \kappa
_2\sigma _2\right) }\right] ^{\left( kk\right) }
|nl^{N}\alpha ^{\prime }
L^{\prime }S^{\prime })- \\
-\left\{
\begin{array}{ccc}
\kappa _1 & \kappa _2 & k \\
l & l & l
\end{array}
\right\} \left\{
\begin{array}{ccc}
\sigma _1 & \sigma _2 & k \\
s & s & s
\end{array}
\right\} \times \\
\left. \times (nl^{N}\alpha LS| \left[ a^{\left( \lambda \right) }\times
\stackrel{\sim }{a}^{\left( \lambda \right) }\right] ^{\left(
kk\right) }  |nl^{N}\alpha L^{\prime }S^{\prime })
\right\}.
\end{array}
\end{equation}

Using the relationships between the tensorial product of creation
and annihilation operators and the tensorial quantities $U^{k}$ and $V^{k1}$
(see Rudzikas and Kaniauskas \cite{ra}), the expression (\ref{eq:ma-f})
for matrix elements can be writen down in terms of $U^{k}$ and $V^{k1}$.
In comparing (\ref{eq:ma-d}) to (\ref{eq:ma-f}) we see that the summation
over intermediate terms
$\alpha_{i}^{\prime \prime} L^{\prime \prime}_{i} S^{\prime \prime}_{i}$,
$\alpha_{i}^{\prime \prime \prime} L^{\prime \prime \prime}_{i}
S^{\prime \prime \prime}_{i}$,
$\alpha_{i}^{\prime V} L^{\prime V}_{i} S^{\prime V}_{i}$ is already
performed in expression (\ref{eq:ma-f}). So, in this case the Racah
algebra is exploited at the level of standard quantities
$U^{k}$ ir $V^{k1}$. This simplifies calculations a lot:
\begin{itemize}
\item For zero matrix elements are easily tracked down from triangular
conditions even before the actual calculation is performed. In case
(\ref{eq:ma-f}) only the triangular conditions
$\delta(L,k,L^{\prime})$ and $\delta(S,k,S^{\prime})$
are present, but their number may be greater in
other cases. (In the above, the notation $\delta(L,k,L^{\prime})$
means the triangular condition $\mid L-L^{\prime } \mid \leq k \leq
L+L^{\prime }$.)

\item The tables of standard quantities (see Nielson and Koster \cite{Nielson},
Karazija {\it et al}~\cite{Kb}, Cowan~\cite{Cowan}) may be used.
\item The recoupling matrix is simpler in this case, and it has an
analytical expression.
\end{itemize}

So, the expressions exploiting the Racah algebra at the level of
$U^{k}$ and $V^{k1}$ are much more advantageous than (\ref{eq:ma-c}).
Such expressions are obtained for all physical operators. For example,
the expressions for spin-other-orbit operator are presented in papers
Horie \cite{Horie}, Karazija {\it et al}~\cite{s-o-a} and
Vizbarait\. e {\it et al}~\cite{s-o-b}, the ones for
spin-spin operator - in papers
Horie \cite{Horie} and Karazija {\it et al}~\cite{s-s}, and the ones
for orbit-orbit operator in the monograph Jucys and Savukynas \cite{js}.
The shortcoming of the expressions of this type is that the Racah algebra
is exploited to its full extent in separate cases only. This is discussed in
detail in paper by Gaigalas {\it et al}~\cite{GR}.

Gaigalas {\it et al} \cite{GRF} have proposed a methodology which allows
one to take all the advantages of the Racah algebra in the most general case.
According to the approach by Gaigalas {\it et al} \cite{GRF},
a general expression of submatrix element
for any two-particle operator between functions with $u$ open shells can be
written down as follows:

\begin{equation}
\label{eq:ma-g}
\begin{array}[b]{c}
(\psi _u^{bra}\left( LS\right) ||
\widehat{G}^{\left( \kappa _1\kappa _2k,\sigma_1\sigma _2k\right) }
||\psi _u^{ket}\left( L^{\prime }S^{\prime
}\right) )= \\
=\displaystyle {\sum_{n_il_i,n_jl_j,n_i^{\prime }l_i^{\prime },n_j^{\prime
}l_j^{\prime }}}
(\psi _u^{bra}\left( LS\right) ||\widehat{G}
\left( n_il_i,n_jl_j,n_i^{\prime }l_i^{\prime },n_j^{\prime }
l_j^{\prime } \right)
||\psi _u^{ket}\left( L^{\prime }S^{\prime
}\right) )= \\
=
\displaystyle {\sum_{n_il_i,n_jl_j,n_i^{\prime }l_i^{\prime },n_j^{\prime
}l_j^{\prime }}}\displaystyle {\sum_{\kappa _{12},\sigma _{12},\kappa
_{12}^{\prime },\sigma _{12}^{\prime }}}\sum \left( -1\right) ^\Delta \Theta
^{\prime }\left( n_i\lambda _i,n_j\lambda _j,n_i^{\prime }\lambda _i^{\prime
},n_j^{\prime }\lambda _j^{\prime },\Xi \right) \times \\ \times T\left(
n_i\lambda _i,n_j\lambda _j,n_i^{\prime }\lambda _i^{\prime },n_j^{\prime
}\lambda _j^{\prime },\Lambda ^{bra},\Lambda ^{ket},\Xi ,\Gamma \right)
\times \\
\times R\left( \lambda _i,\lambda _j,\lambda _i^{\prime },\lambda _j^{\prime},
\Lambda ^{bra},\Lambda ^{ket},\Gamma \right).
\end{array}
\end{equation}

In calculating the spin-angular part of a submatrix element using
(\ref{eq:ma-g}), one has to compute the following quantities
(for more detail see Gaigalas~\cite{GRF}):

\begin{enumerate}
\item  The recoupling matrix $R\left( \lambda _i,\lambda _j,\lambda
_i^{\prime },\lambda _j^{\prime },\Lambda ^{bra},\Lambda ^{ket},\Gamma
\right) $. This recoupling matrix accounts for the change in
going from matrix element

$(\psi _u^{bra}\left( LS\right) ||
G||\psi _u^{ket}\left( L^{\prime }S^{\prime }\right) )$,
which has $u$ open shells in the bra and ket functions, to the submatrix
element

$T\left(
n_i\lambda _i,n_j\lambda _j,n_i^{\prime }\lambda _i^{\prime },n_j^{\prime
}\lambda _j^{\prime },\Lambda ^{bra},\Lambda ^{ket},\Xi ,\Gamma \right) $,
which has only the shells being  acted upon by the two-particle operator in its bra and ket
functions.

\item
The submatrix element
$T\left( n_i\lambda _i,n_j\lambda _j,n_i^{\prime }\lambda _i^{\prime
},n_j^{\prime }\lambda _j^{\prime },\Lambda ^{bra},\Lambda ^{ket},\Xi
,\Gamma \right) $, which
denotes
the submatrix elements of operators (\ref{eq:op-j}) - (\ref{eq:op-n}).
Here $\Gamma $ refers to the array of coupling parameters
connecting the recoupling matrix $R\left( \lambda _i,\lambda _j,\lambda
_i^{\prime },\lambda _j^{\prime },\Lambda ^{bra},\Lambda ^{ket},\Gamma
\right) $ to the submatrix element.

\item  Phase factor $\Delta $ (for more detail see Gaigalas~\cite{GRF}).

\item  $\Theta ^{\prime }\left( n_i\lambda _i,n_j\lambda _j,n_i^{\prime
}\lambda _i^{\prime },n_j^{\prime }\lambda _j^{\prime },\Xi \right) $, which
is proportional to the radial part and corresponds to one of
$\Theta \left(
n\lambda ,\Xi \right) $,...,$\Theta \left( n_\alpha \lambda _\alpha ,n_\beta
\lambda _\beta ,n_\gamma \lambda _\gamma ,n_\delta \lambda _\delta ,\Xi
\right) $. It consists of a submatrix element
 $\left( n_i\lambda
_in_j\lambda _j||g^{\left( \kappa _1\kappa _2k,\sigma _1\sigma _2k\right)
}||n_i^{\prime }\lambda _i^{\prime }n_j^{\prime }\lambda _j^{\prime }\right)
$, and in some cases of simple factors and 3$nj$-coefficients
(for more detail see Gaigalas~\cite{GRF}).
\end{enumerate}

In the next section we shall discuss finding of the recoupling matrix.
Now we shall analyse the submatrix element
$T\left( n_i\lambda _i,n_j\lambda _j,n_i^{\prime }\lambda _i^{\prime
},n_j^{\prime }\lambda _j^{\prime },\Lambda ^{bra},\Lambda ^{ket},\Xi
,\Gamma \right) $. As the angular part of expression (\ref{eq:ma-g})
contains the tensors
(\ref{eq:op-j}) - (\ref{eq:op-n}), so
we will discuss the derivation of submatrix elements of
these operators, and present the expressions for
these quantities. It is worth noting that these tensorial quantities all act
upon the {\em same} shell. So, all the advantages of tensor algebra and the
quasispin formalism may be exploited efficiently.

We obtain the submatrix elements of operator (\ref{eq:op-j}) by
straightforwardly using the Wigner-Eckart theorem in quasispin space
(see Rudzikas \cite{r}):
\begin{equation}
\label{eq:ma-h}
\begin{array}[b]{c}
\left( l^N\;\alpha QLS||a_{m_q}^{\left( qls\right) }||l^{N^{\prime
}}\;\alpha ^{\prime }Q^{\prime }L^{\prime }S^{\prime }\right) =-\left[
Q\right] ^{-1/2}\left[
\begin{array}{ccc}
Q^{\prime } & 1/2 & Q \\
M_Q^{\prime } & m_q & M_Q
\end{array}
\right] \times \\
\times \left( l\;\alpha QLS|||a^{\left( qls\right) }|||l\;\alpha ^{\prime
}Q^{\prime }L^{\prime }S^{\prime }\right) ,
\end{array}
\end{equation}
where the last multiplier in (\ref{eq:ma-h}) is the so-called completely
reduced (reduced in the quasispin, orbital and spin spaces) matrix element.

The value of the submatrix element of operator (\ref{eq:op-k}) is obtained by

\begin{equation}
\label{eq:ma-i}
\begin{array}[b]{c}
\left( nl^N\;\alpha QLS||\left[ a_{m_{q1}}^{\left( q\lambda \right) }\times
a_{m_{q2}}^{\left( q\lambda \right) }\right] ^{\left( k_1k_2\right)
}||nl^{N^{\prime }}\;\alpha ^{\prime }Q^{\prime }L^{\prime }S^{\prime
}\right) = \\
=\displaystyle {\sum_{\epsilon ,m_\epsilon }}\left[ Q\right] ^{-1/2}\left[
\begin{array}{ccc}
q & q & \epsilon \\
m_{q1} & m_{q2} & m_\epsilon
\end{array}
\right] \left[
\begin{array}{ccc}
Q^{\prime } & \epsilon & Q \\
M_Q^{\prime } & m_\epsilon & M_Q
\end{array}
\right] \times \\
\times \left( nl\;\alpha QLS|||W^{\left( \epsilon k_1k_2\right)
}|||nl\;\alpha ^{\prime }Q^{\prime }L^{\prime }S^{\prime }\right) .
\end{array}
\end{equation}

On the right-hand side of equations (\ref{eq:ma-h}) and (\ref{eq:ma-i})
only the Clebsch-Gordan coefficient $\left[
\begin{array}{ccc}
Q^{\prime } & \epsilon & Q \\
M_Q^{\prime } & m_\epsilon & M_Q
\end{array}
\right] $ depends on the number $N$ of equivalent electrons.

$\left( nl\;\alpha QLS|||W^{\left( \epsilon k_1k_2\right) }|||nl\;\alpha
^{\prime }Q^{\prime }L^{\prime }S^{\prime }\right) $ denotes reduced in
quasispin space submatrix element (completely reduced matrix element) of the
triple tensor $W^{\left( \epsilon k_1k_2\right) }\left( nl,nl\right) =\left[
a^{\left( qls\right) }\times a^{\left( qls\right) }\right] ^{\left( \epsilon
k_1k_2\right) }$. It is related to the completely reduced coefficients
(subcoefficients) of fractional parentage in a following way:

\begin{equation}
\label{eq:ma-j}
\begin{array}[b]{c}
\left( nl\;\alpha QLS|||W^{\left( \epsilon k_1k_2\right) }|||nl\;\alpha
^{\prime }Q^{\prime }L^{\prime }S^{\prime }\right) = \\
=\left( -1\right) ^{Q+L+S+Q^{\prime }+L^{\prime }+S^{\prime }+\epsilon
+k_1+k_2}\left[ \epsilon ,k_1,k_2\right] ^{1/2}\times \\
\times
\displaystyle {\sum_{\alpha ^{\prime \prime }Q^{\prime \prime }L^{\prime
\prime }S^{\prime \prime }}}\left( l\;\alpha QLS|||a^{\left( qls\right)
}|||l\;\alpha ^{\prime \prime }Q^{\prime \prime }L^{\prime \prime }S^{\prime
\prime }\right) \times \\ \times \left( l\;\alpha ^{\prime \prime }Q^{\prime
\prime }L^{\prime \prime }S^{\prime \prime }|||a^{\left( qls\right)
}|||l\;\alpha ^{\prime }Q^{\prime }L^{\prime }S^{\prime }\right) \times \\
\times \left\{
\begin{array}{ccc}
q & q & \epsilon \\
Q^{\prime } & Q & Q^{\prime \prime }
\end{array}
\right\} \left\{
\begin{array}{ccc}
l & l & k_1 \\
L^{\prime } & L & L^{\prime \prime }
\end{array}
\right\} \left\{
\begin{array}{ccc}
s & s & k_2 \\
S^{\prime } & S & S^{\prime \prime }
\end{array}
\right\} .
\end{array}
\end{equation}

In the other three cases (\ref{eq:op-l}), (\ref{eq:op-m}), (\ref{eq:op-n})
we obtain the submatrix elements of these operators
by using (2.28) of Jucys and Savukynas \cite{js}:
\begin{equation}
\label{eq:ma-k}
\begin{array}[b]{c}
(nl^N\;\alpha QLS||\left[ F^{\left( \kappa _1\sigma _1\right) }\left(
n\lambda \right) \times G^{(\kappa _2\sigma _2)}\left( n\lambda \right)
\right] ^{\left( kk\right) }||nl^{N^{\prime }}\;\alpha ^{\prime }Q^{\prime
}L^{\prime }S^{\prime })= \\
=\left( -1\right) ^{L+S+L^{\prime }+S^{\prime }+2k}\left[ k\right] \times \\
\times
\displaystyle {\sum_{\alpha ^{\prime \prime }Q^{\prime \prime }L^{\prime
\prime }S^{\prime \prime }}}(nl^N\;\alpha QLS||F^{\left( \kappa _1\sigma
_1\right) }\left( n\lambda \right) ||nl^{N^{\prime \prime }}\;\alpha
^{\prime \prime }Q^{\prime \prime }L^{\prime \prime }S^{\prime \prime
})\times \\ \times (nl^{N^{\prime \prime }}\;\alpha ^{\prime \prime
}Q^{\prime \prime }L^{\prime \prime }S^{\prime \prime }||G^{(\kappa _2\sigma
_2)}\left( n\lambda \right) ||nl^{N^{\prime }}\;\alpha ^{\prime }Q^{\prime
}L^{\prime }S^{\prime })\times \\
\times \left\{
\begin{array}{ccc}
\kappa _1 & \kappa _2 & k \\
L^{\prime } & L & L^{\prime \prime }
\end{array}
\right\} \left\{
\begin{array}{ccc}
\sigma _1 & \sigma _2 & k \\
S^{\prime } & S & S^{\prime \prime }
\end{array}
\right\} ,
\end{array}
\end{equation}
where $F^{\left( \kappa _1\sigma _1\right) }\left( n\lambda \right) $,
$G^{(\kappa _2\sigma _2)}\left( n\lambda \right) $ are one of (\ref{eq:op-j})
or
(\ref{eq:op-k}) and the submatrix elements correspondingly are defined by (\ref
{eq:ma-h}), (\ref{eq:ma-i}) and (\ref{eq:ma-j}). $N^{\prime \prime }$
is defined by the second quantization operators occurring in $F^{\left(
\kappa _1\sigma _1\right) }\left( n\lambda \right) $ and $G^{(\kappa
_2\sigma _2)}\left( n\lambda \right) $.

As is seen, by using this approach  Gaigalas {\it et al} \cite{GRFb},
the calculation of the angular parts of
matrix elements between functions with $u$ open shells is
reduced to requiring the reduced coefficients of fractional parentage or
the tensors (for example $W^{\left( \epsilon k_1k_2\right) }\left(
nl,nl\right) $), which are independent of the occupation number of the shell
and are acting on one shell of equivalent electrons.

The main advantage of this approach is that the standard
data tables in such a case will be much smaller in comparison with tables of
the usual coefficients $U^k,$ $V^{k_1k_2}$
(see Jucys and Savukynas \cite{js})
and, therefore, many summations will be less
time-consuming. Also one can see that in such an approach the submatrix
elements of standard tensors and subcoefficients of fractional parentage
actually can be treated in a uniform way as they all are the completely
reduced matrix elements of the second quantization operators. Hence, all
methodology of calculation of matrix elements will be much more universal in
comparison with the traditional one (see Cowan~\cite{Cowan},
Jucys and Savukynas \cite{js}, Wybourne \cite{Wybourne}).

\section{ Recoupling Matrix }

While seeking the matrix elements of one- or two- particle operators,
it is necessary to obtain the values of a recoupling matrix

$R\left( \lambda _i,\lambda _j,\lambda
_i^{\prime },\lambda _j^{\prime },\Lambda ^{bra},\Lambda ^{ket},\Gamma
\right) $, if we use the methodology by
Gaigalas {\it et al}~\cite{GRF} (see expresion (\ref{eq:ma-g})), or the
recoupling matrices

$R_{d} \left( \lambda _i,\lambda _j,\lambda _i^{\prime },\lambda _j^{\prime},
\kappa_1, \kappa_2, \kappa, \sigma_1, \sigma_2, \sigma, \Lambda ^{bra},
\Lambda ^{ket} \right)$
and

$R_{e} \left( \lambda _i,\lambda _j,\lambda _i^{\prime },\lambda _j^{\prime},
\kappa_1, \kappa_2, \kappa, \sigma_1, \sigma_2, \sigma, \Lambda ^{bra},
\Lambda ^{ket} \right)$, if we use the methodology by Fano~\cite{Fano}
(see expresion (\ref{eq:ma-c})). In the case of several open shells the
expressions for matrix elements of every physical operator are published,
where the recoupling matrices are in the form of simple factors. Usually all
these are written in the coordinate representation. They can be found in
Karazija {\it et al}~\cite{Kb},
Jucys and Savukynas~\cite{js}, Karazija~\cite{Ka}, Rudzikas~\cite{ra}.

Meanwhile, for the more complex configurations, i.e. the ones having many
shells, the recoupling matrices are much more complicated. Beside that, the
complexity of a two-particle operator adds to this. When the tensorial
structure of an operator is complex, the recoupling matrix is rather complex,
too, e.g. the spin-other-orbit operator (see Gaigalas~\cite{GBRF}).
While attempting to calculate the angular part of matrix elements in all the
mentioned cases, a general methodology for calculating the recoupling
matrices is necessary. It has to be efficient, too, because the speed of
calculation of angular parts of matrix elements depends on that.

The majority of methodologies to obtain angular parts are based on the
Fano~\cite{Fano} calculation scheme (see for example Glass~\cite{Glass},
Glass and Hibbert~\cite{GH}, Grant~\cite{Grant}). In finding the matrix
elements using this, one of the tasks is to obtain the recoupling matrices for
direct and exchange terms.

\begin{figure}
\centering
\includegraphics[height=13cm]{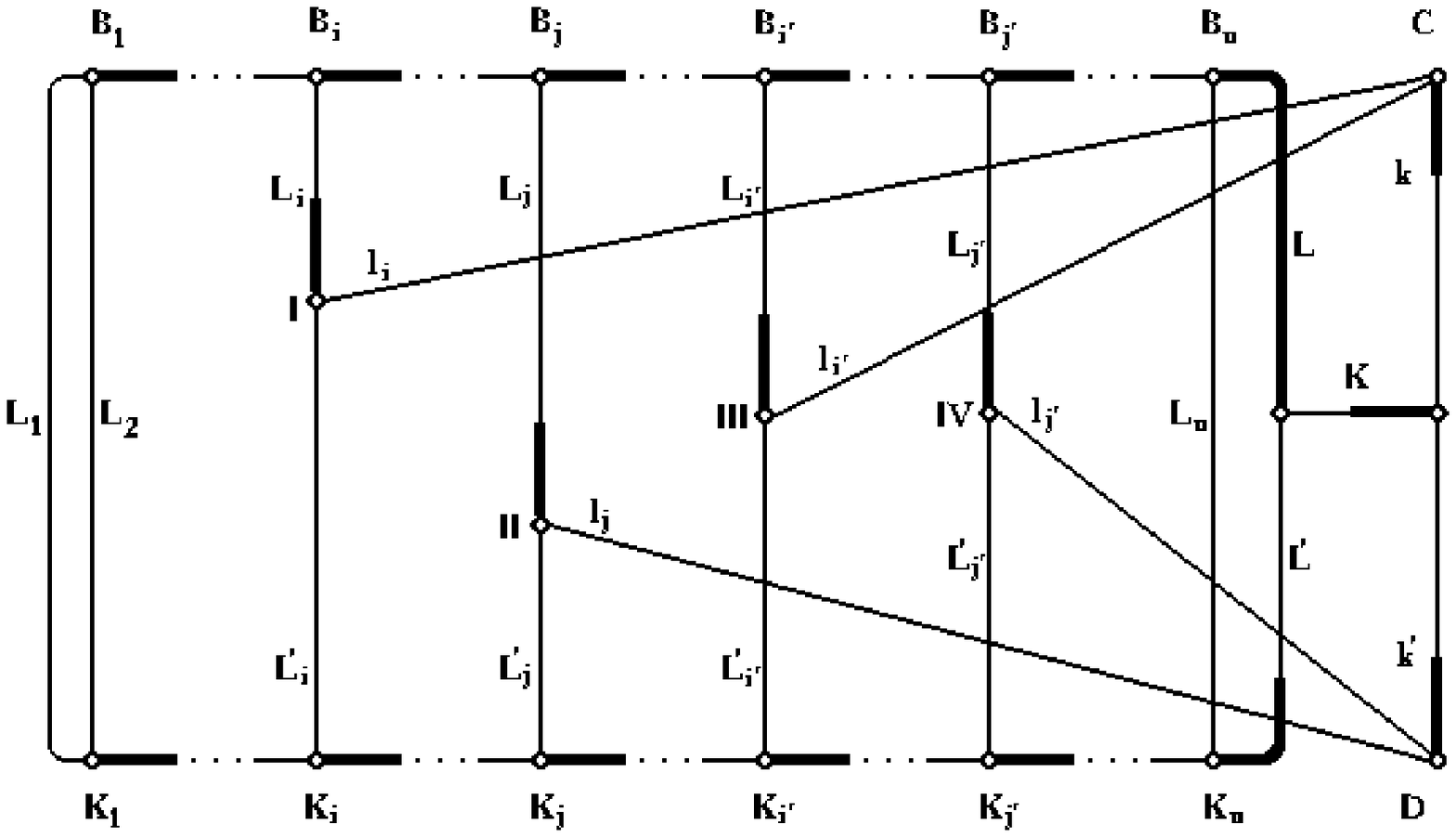}
\vspace{-5.cm}
\caption{The recoupling matrix for the Fano angular calculation scheme.}
\label{fig:a}
\end{figure}

Let us treat a matrix element of a two-particle operator acting upon
four distinct shells

\begin{equation}
\label{eq:r-a}
\begin{array}{c}
\left( \psi _u^{bra}\left( LS M_LM_S \right)||\widehat{G}(ij,i'j')||
\psi _u^{ket}\left( L'S' M'_LM'_S \right) \right).
\end{array}
\end{equation}
For such an operator, the recoupling matrix in $L$ - space, using
graphical technique of Jucys and Bandzaitis~\cite{jb}, is represented in
Figure~\ref{fig:a}. As we speak of calculations in $LS$ - coupling, the
analogous recoupling matrix in $S$ - space has to be calculated.
In the Figure~\ref{fig:a} the tree containing {\bf B} nodes represents the
bra function, and the one with {\bf K} nodes represents the ket function.
Bra, as well as ket function, both contain $u$ open shells in this case.

Now let us treat a case when three distinct shells are acted upon by the
operator. For example, the matrix element of operator $\widehat{G}(ij,ij')$
is sought. Then in the recoupling matrix the node III will be on $L_i$
line (see Figure~\ref{fig:a}). This means that there will be two nodes on the
resulting orbital moment $L_i$ line of shell $i$. These will be nodes I and III.
The resulting orbital moment $L_{i}$ of the shell $i$ of bra function is
attributed to the line connecting nodes $B_i$ and I. The resulting orbital
moment $L'_i$ of the shell $i$ of ket function is attributed to the line
between nodes III and $K_i$. Finally, characteristics $L''_i$ of all
possible intermediate momenta for that particular shell, which are summed over,
are attributed to the line between nodes I and III. So, in treating the
operator acting upon three distinct shells, an intermediate sum in the
recoupling matrix appears. Similarly, two intermediate sums occur when two
distinct shells are acted upon by the operator. And when it acts upon one
shell only, three intermediate sums are present. The lines $k$, $k'$ and $K$
in the recoupling matrix represent the structure of two- particle operator
in $l$ - space. See Tutlys~\cite{Tutlys}, Grant~\cite{Grant} for details.

This is also valid for the exchange term, only the line $l_{i'}$ must
be connected to the node {\bf D}, and line $l_{j'}$ - to the node {\bf C}.

The first program to calculate the recoupling matrices of this type, NJSYM,
was written by Burke~\cite{Burke}. It performs the calculations in two stages:
1) the recoupling matrix is expressed as a sum of products of the $6j$ -
coefficients; 2) this expression is used in calculation.

Tutlys~\cite{Tutlys} wrote a program to calculate angular parts of matrix
elements, ANGULA, which expressed the recoupling matrix in terms of
Clebsch - Gordan coefficients before the actual calculations. While finding
the recoupling matrix by the Clebsch - Gordan coefficient summation, this
program eliminates trivial coefficients from  the expression.

Bar-Shalom and Klapisch~\cite{BSK} developed a new program NJGRAF.
This program calculates the recoupling matrix in several stages. On the basis
of graphical methodology by Yutsis, Levinson and Vanagas~\cite{YLV}, the
recoupling matrix is analysed graphically and an optimal expression is found.
Afterwards, the value of recoupling matrix itself is calculated. An analogous
program RECOUP was written by Lima~\cite{Lima}, and a program NEWGRAPH
was written by Fack {\it et al}~\cite{FPV}. All these (NJGRAF, RECOUP and
NEWGRAPH) are based on the same principle. An optimal analytical expression
for the recoupling matrix is obtained by graphical method, and then the
calculations are carried out according to it. But the
optimal expressions they find are different quite often, and are not really
optimal.

As it was mentioned above, the methodology of angular calculation based on the
Fano calculation scheme has a shortcoming that the intermediate sums appear
in complex recoupling matrices. Due to these summations
and the complexity of the recoupling matrix itself, the associated computer
codes become rather time consuming. A solution to this problem was found by
Burke {\it et al}~\cite{BBD}. They tabulated separate standard parts of
recoupling matrices along with coefficients of fractional parentage at the
beginning of a calculation and then used them later to calculate the
coefficients needed.

Computer codes by Glass~\cite{Glass},
Glass and Hibbert~\cite{GH}, Burke {\it et al}~\cite{BBD}, Fischer~\cite{fb},
Fischer~\cite{fc} and Dyall {\it et al}~\cite{GRASP}
utilize the program NJSYM (Burke~\cite{Burke}) or NJGRAF
(Bar-Shalom and Klapisch~\cite{BSK}) for the calculation of recoupling matrices.
Both are rather time consuming when calculating matrix elements of complex
operators or electronic configurations when calculating matrix elements of
complex operators or electronic configurations with many open shells.
In order to simplify the calculations, Cowan~\cite{Cowan} suggested grouping
matrix elements into 'classes' (see Cowan~\cite{Cowan}, Figure 13-15).
Unfortunately this approach was not generalized to all two-electron operators.
Perhaps this is the reason why Cowan's approach is not widely used although
the program itself, based on this approach is widely used.

\begin{figure}
\centering
\includegraphics[height=13.5cm]{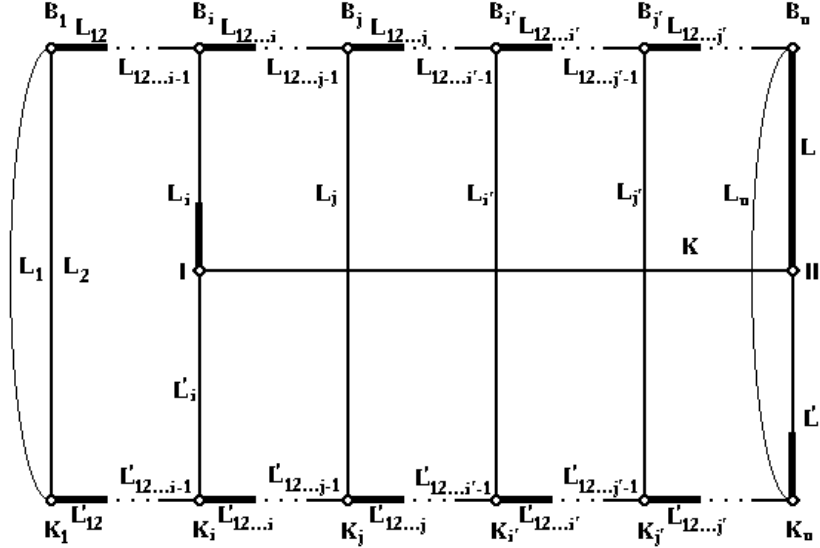}
\vspace{-6.cm}
\caption{Diagram $R_{1}$ representing the recoupling matrix
$R\left( \lambda _i,\lambda _j,\lambda
_i^{\prime },\lambda _j^{\prime },\Lambda ^{bra},\Lambda ^{ket},\Gamma
\right) $ when a two-particle operator acts upon a single shell.}
\label{fig:b}
\end{figure}

\begin{figure}
\centering
\includegraphics[height=13.5cm]{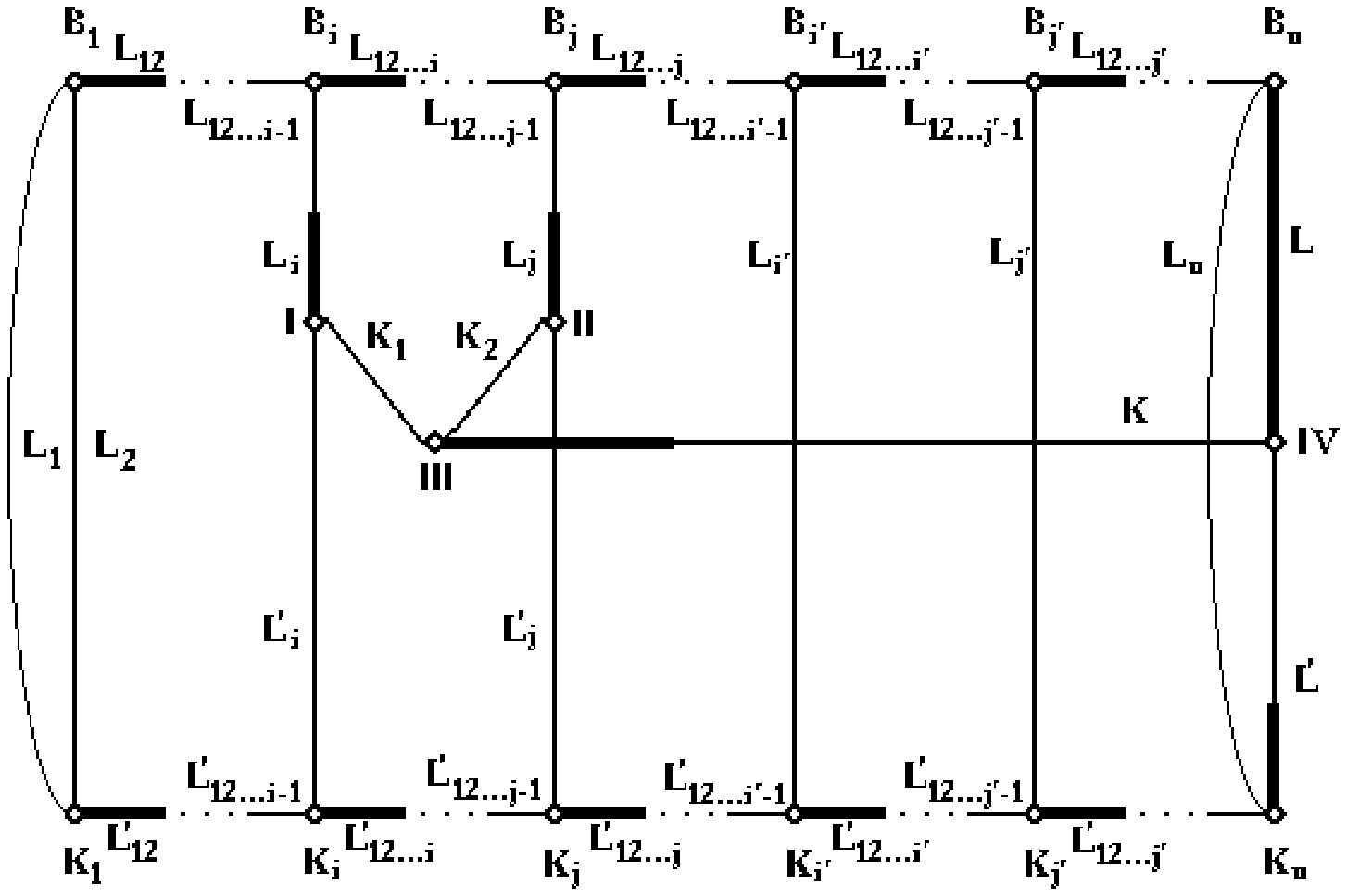}
\vspace{-6.cm}
\caption{Diagram $R_{2}$ representing the recoupling matrix
$R\left( \lambda _i,\lambda _j,\lambda
_i^{\prime },\lambda _j^{\prime },\Lambda ^{bra},\Lambda ^{ket},\Gamma
\right) $ when a two-particle operator acts upon a two shell.}
\label{fig:c}
\end{figure}

\begin{figure}
\centering
\includegraphics[height=13.5cm]{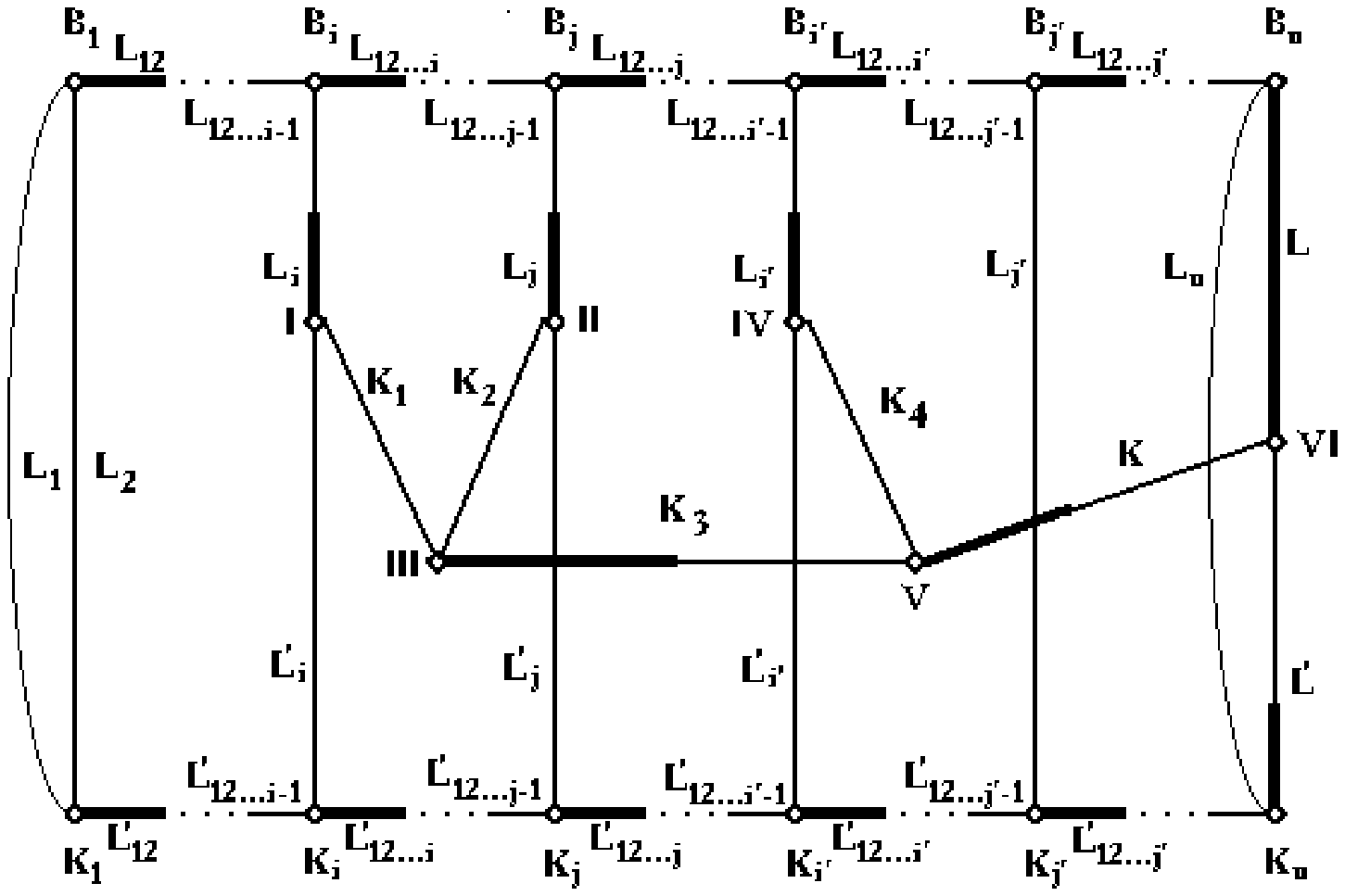}
\vspace{-6.cm}
\caption{Diagram $R_{3}$ representing the recoupling matrix
$R\left( \lambda _i,\lambda _j,\lambda
_i^{\prime },\lambda _j^{\prime },\Lambda ^{bra},\Lambda ^{ket},\Gamma
\right) $ when a two-particle operator acts upon a three shell.}
\label{fig:d}
\end{figure}

\begin{figure}
\centering
\includegraphics[height=13.5cm]{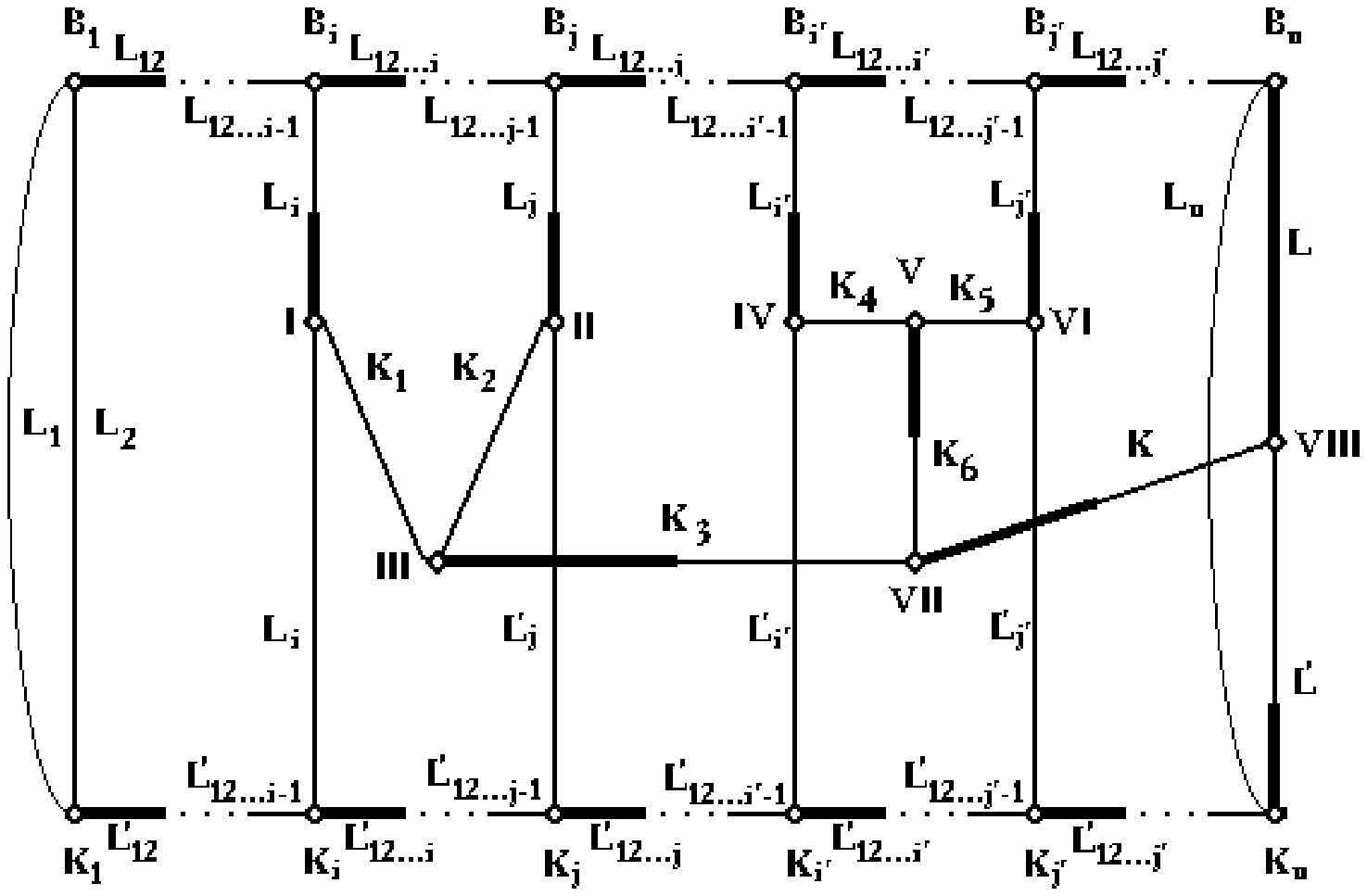}
\vspace{-6.cm}
\caption{Diagram $R_{4}$ representing the recoupling matrix
$R\left( \lambda _i,\lambda _j,\lambda
_i^{\prime },\lambda _j^{\prime },\Lambda ^{bra},\Lambda ^{ket},\Gamma
\right) $ when a two-particle operator acts upon a four shells.}
\label{fig:e}
\end{figure}

Gaigalas {\it et al}~\cite{GRF} proposed a methodology where the analytical
expressions for recoupling matrices are obtained for the most general case.
In this methodology, analogically as in Cowan~\cite{Cowan}, the matrix elements
are attributed to four different groups. The operators acting upon only one
shell belong to the first group (distribution 1 from Table 1),
the ones acting upon two - to the second (distributions 2 - 10 from Table 1),
upon three - to the third (distributions 11 - 18 from Table 1),
and upon four - to the fourth group (distributions 19 - 42 from Table 1)
respectively.
Each group has a different recoupling matrix, $R_{1}$ - $R_{2}$. They all are
shown in Figures~\ref{fig:b} - \ref{fig:e}.

\subsection{ One interacting shell}

Let us treat the recoupling matrix $R_1$ first (see Figure~\ref{fig:b}).
It represents the case when a physical two-particle operator acts upon a
single
shell. Here, similarly as in Figure~\ref{fig:a}, the graphical technique of
Jucys and Bandzaitis~\cite{jb} is used. The bra and ket functions are
represented as in Figure~\ref{fig:a}. As the two-particle operator acts upon
only one shell in this case, this operator may be treated as a single -
particle one from the recoupling matrix point of view. Using a graphical rule
allowing one to cut two lines and connect the loose ends, we disconnect the
nodes $B_1$ and $K_1$ from the general recoupling matrix. The part separated
from recoupling corresponds to delta - functions only,
$\delta (L_{12}, L'_{12})$, $\delta(L_1,L_2,L_{12})$.
In addition, using graphical technique it is possible to cut all the nodes
until $B_i$ and $K'_{i}$ out of the general recoupling matrix. All the cut
nodes contribute the same delta functions which may be written as
$\delta (L_{12..i-1}, L'_{12..i-1})$, $\delta(L_{12..i-1},L_i,L_{12..i+1})$.

\begin{figure}
\centering
\includegraphics[height=14cm]{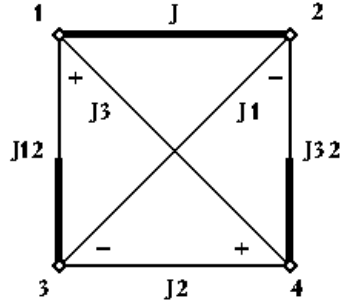}
\vspace{-10.cm}
\caption{Diagram corresponding to the notation
$A_{1}(+ - + - ; J_{1} J_{2} J_{12} ; J_{3} J J_{32} )$.}
\label{fig:f}
\end{figure}

In the next stage, it remains to treat the part of recoupling matrix from the
nodes $B_i$ and $F_i$ up to $B_u$ and $K_u$. Beside that, the lines
$L_{12..i-1}$ and $L'_{12..i-1}$ are connected in this remaining diagram.
Before obtaining the analytical expression for the remaining recoupling
matrix we introduce a notation
$A_{1}(+ - + - ; J_{1} J_{2} J_{12} ; J_{3} J J_{32} )$.
It will help us to describe the procedure of obtaining analytical expression
for the remaining recoupling matrix in a simpler way. So, the diagram in
Figure~\ref{fig:f} is denoted as
$A_{1}(+ - + - ; J_{1} J_{2} J_{12} ; J_{3} J J_{32} )$.
It equals to

\begin{equation}
\label{eq:r-b}
A_{1}(+ - + - ; J_{1} J_{2} J_{12} ; J_{3} J J_{32} )=
\left( -1\right) ^{2j} \left[ j_{12}, j_{32} \right]
^{1/2}\left\{
\begin{array}{ccc}
J_{1} & J_{2} & J_{12} \\
J_{3} & J     & J_{32}
\end{array}
\right\} .
\end{equation}

Now, using the graphical rule of cutting three lines, we cut out the nodes
$I$, $B_{i}$ and $K_{i}$ from the recoupling matrix. Thus we obtain a diagram
$A_1(- - + + ; K L'_{i} L_{i} ; L_{12..i-1} L L'_{12..i} )$. This coefficient
in the paper by Gaigalas {\it et al}~\cite{GRF} is denoted as $C_{1}$, and
equals to
\begin{equation}
\label{eq:r-c}
C_1=\left( -1\right) ^\varphi \left[ L_a,T^{\prime }\right]
^{1/2}\left\{
\begin{array}{ccc}
k & L_a^{\prime } & L_a \\
J & T & T^{\prime }
\end{array}
\right\} .
\end{equation}

The diagram is cut from the nodes $B_{j}$ and $K_{j}$ up to
$B_{u}$ and $K_{u}$ in the same way. It is expressed in terms of diagrams

\begin{equation}
\label{eq:r-d}
\begin{array}[b]{c}
C_2\left( K,k_{\min },k_{\max }\right) =
\displaystyle {\stackrel{k_{\max }}{\prod_{i=k_{\min }}}}
A_{1}(+ - + -; K L'_{12...i} L_{12...i} ; L_{i} L_{12...i-1} L'_{12...i-1}).
\end{array}
\end{equation}

It is easy to notice that $C_2\left( K,k_{\min },k_{\max }\right)$
corresponds to the coefficient $C_2\left( K,k_{\min },k_{\max }\right)$
defined in Gaigalas {\it et al}~\cite{GRF}.

\subsection{ Two interacting shells}

\begin{figure}
\centering
\includegraphics[height=12cm]{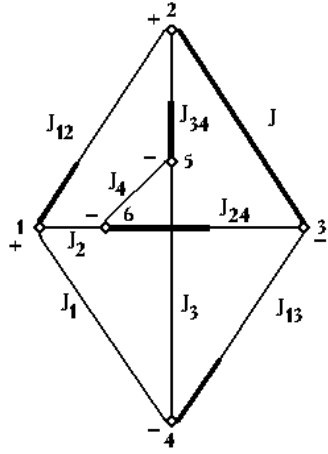}
\vspace{-5.cm}
\caption{Diagram corresponding to the notation
$A_{2}(- + + - - -; J_{1} J_{2} J_{12} ; J_{3} J_{4} J_{34}; J_{13} J_{24} J )$.}
\label{fig:g}
\end{figure}

Before going into investigation of the recoupling matrix pictured in
Figure~\ref{fig:c}, we denote the diagram $A_{2}$ (see
Figure~\ref{fig:g} ) as

$A_{2}(- + + - - -; J_{1} J_{2} J_{12} ; J_{3} J_{4} J_{34}; J_{13} J_{24} J )$.
Its analytical expression is

\begin{equation}
\label{eq:r-e}
A_{2}(- + + - - -; J_{1} J_{2} J_{12} ; J_{3} J_{4} J_{34}; J_{13} J_{24} J )
=\left[ J_{12},J_{34},J_{13},J_{24} \right] ^{1/2}\left\{
\begin{array}{ccc}
J_{1}  & J_{2}  & J_{12} \\
J_{3}  & J_{4}  & J_{34} \\
J_{13} & J_{24} & J
\end{array}
\right\} .
\end{equation}
In the $A_2$ diagram, first the signs of nodes 1-6, and then the moments
are presented.

While investigating the recoupling matrix (see Figure~\ref{fig:c}) it is
easy to notice that there is an additional graphic element. After cutting this
out, we get a diagram the further investigation of which is very similar to
the investigation of $R_{1}$. That additional matrix element is

$A_{2}(- + + - - -; L'_{12..j-1} K_{1} L_{12..j-1}; J'_{j} K_{2} L_{j};
L'_{12..j} K L_{12..j})$.
It is obtained by cutting the nodes $B_{j}$, $II$, $III$, $K_{j}$ out of the
diagram $R_2$. After doing this, the diagram under investigation is split
into two diagrams closely resembling the recoupling matrix $R_1$. As they are
cut in the very same way as $R_1$ is, we will not stop at recoupling matrix
$R_2$ for details.

\subsection{ Three interacting shells}

In the methodology presented in Gaigalas {\it et al}~\cite{GRF}, the
recoupling matrix is represented by a diagram $R_3$ (see Figure~\ref{fig:d})
when the two-particle operator is acting upon three shells. Similarly
as for the recoupling matrix $R_2$, the cutting out of nodes $B_{j}$, $II$,
$III$ and $K_j$ leads to a diagram
$A_{2}(- + + - - -; L'_{12..j-1} K_{1} L_{12..j-1}; J'_{j} K_{2} L_{j};
L'_{12..j} K_3 L_{12..j})$, and the cutting out of nodes
$B'_{i}$, $IV$, $V$ ir $K'_i$ leads to a diagram

$A_{2}(- + + - - -; L'_{12..i'-1} K_{3} L_{12..i'-1}; J'_{i'} K_{4} L_{i'};
L'_{12..i'} K L_{12..i'})$. In that case the diagram splits up into three
simple diagrams, the further investigation of which is analogous to that of
$R_1$ recoupling matrix.

\subsection{ Four interacting shells}

\begin{figure}
\centering
\includegraphics[height=13.5cm]{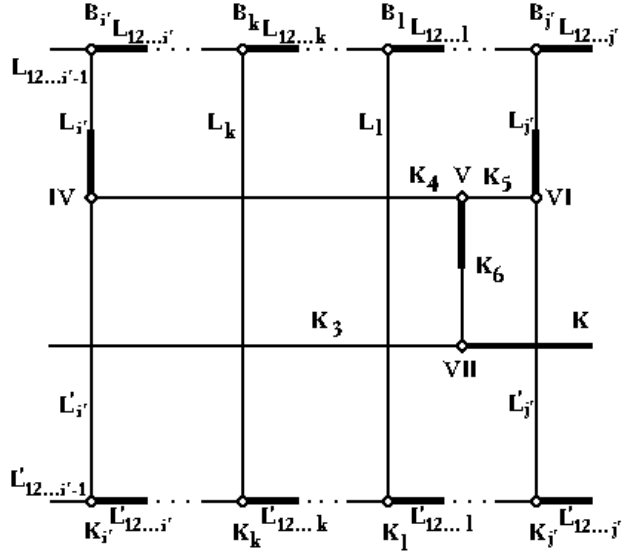}
\vspace{-6.cm}
\caption{The most complex part of the recoupling matrix $R_4$.}
\label{fig:e-b}
\end{figure}

The most complex recoupling matrix $R_4$ (see Figure~\ref{fig:e})
comes into view when all the second quantization operators act upon different
shells. The part of this recoupling matrix from nodes $B_1$ and $K_1$ to nodes
$B_{i'}$ and $K_{i'}$, and also from nodes $B_{j'}$ and $K_{j'}$  to nodes
$B_u$ and $K_u$ is very similar to the recoupling matrices described above,
therefore their investigation is the same as that of the diagrams described
earlier. We will investigate in more  detail the part of diagram from nodes
$B_{i'}$ and $K_{i'}$ to nodes $B_{j'}$ and $K_{j'}$. That is pictured in
Figure~\ref{fig:e-b}. After cutting the diagram across four lines
$L_{12...i'}$, $K_{4}$, $K_{3}$ and $L'_{12...i'}$ and connecting these
according the graphical technique of Jucys and Bandzaitis~\cite{jb}
with a generalized Clebsch-Gordan coefficient, then connecting the lines
$L_{12...i'-1}$, $K_{3}$ and $L'_{12...i'-1}$ with an ordinary
Clebsch-Gordan coefficient, we obtain a diagram which is expressed via two
$6j$-coefficients. The exact analytical expression of this diagram is
presented in the paper by Gaigalas {\it et al}~\cite{GRF} (see expresion (29)).

After cutting the nodes $B_{j'}$, $V$, $VI$, $VII$ and $K_{j'}$ out of
the remaining part and investigating the obtained diagram according to the
Jucys and Bandzaitis~\cite{jb} technique, we obtain a diagram which
corresponds to a $12j$- coefficient. The analytical expression of that is
presented in paper by
Gaigalas {\it et al}~\cite{GRF}, expr. (31).

If a shell $n_{j'}l_{j'}$ is beside shell $n_{i'}l_{i'}$, i.e. there are no
additional nodes between nodes $B_{i'}$, $K_{i'}$ and $B_{j'}$, $K_{j'}$,
then the additional graphical elements vanish. So, we obtained the full
analytical expression of recoupling matrix $R_{4}$. If between the
abovementioned shells a single additional shell is present, then we
obtain an additional diagram which is expressed as a product of two
$6j$-coefficients (see (30) from Gaigalas {\it et al}~\cite{GRF}).
If the number of shells is greater, then the obatined diagram in the most
general case is written via (32) from Gaigalas {\it et al}~\cite{GRF}.

\section{Conclusion}

The methodology that is based on the second quantization in coupled tensorial
form, on the angular momentum theory in three spaces (orbital, spin and
quasispin), on the Wick's theorem
and on the generalized graphical technique of angular momentum,
give the possibility to efficiently calculate the
matrix elements of energy operators
in general case
(Gaigalas and Rudzikas~\cite{GR}, Gaigalas {\it et al}~\cite{GRF}, this paper).
The main ideas of this methodology are:

\begin{enumerate}
\item A number of theoretical methods is known in atomic physics that
facilitate a lot the treatment of angular part of matrix elements.
These are the theory of angular momentum, its graphical representation,
the quasispin, the second quantization and its coupled tensorial form.
But while treating the matrix elements of physical operators in general,
the methods mentiond above are applied only some part, or very inefficiently.

An idea is presented and carried out in the work
Gaigalas and Rudzikas~\cite{GR}, Gaigalas {\it et al}~\cite{GRF} and this paper,
of unifying all these methods in order to optimize the way to treat
matrix elements of physical operators, to investigate efficiently
even the most complex cases of atoms and ions.

\item The Wick's theorem is widely known in the theory of atom. Up to
now it was applied to the most general products of the operators of
second quantization, i.e. when the particular arrays of quantum numbers
for each operator of second quantization were not yet defined. In the
methodology described there is proposed
to apply the Wick's theorem for the products of second quantization
operators where they have the values of quantum numbers already defined. This
allows one, using the methodology of second quantization in coupled tensorial
form, to abtain immediately the optimal tensorial expressions for any operator.
Then, in treting the matrix elements of physical operators, the advantages
of a new modification of the Racah algebra are exploited to their full extent.

\item While analysing the physical operators presented as products of tensors
$a^{(qsl)}$, $W^{(k_q k_l k_s)}$,
$[a^{(qsl)} \times W^{(k_q k_l k_s)}]^{(K_q K_l K_s)}$,
$[W^{(k_q k_l k_s)} \times a^{(qsl)}]^{(K_q K_l K_s)}$,

$[W^{(k_q k_l k_s)} \times W^{(k'_q k'_l k'_s)}]^{(K_q K_l K_s)}$,
it is possible to obtain convenient analytical expressions for recoupling
matrices that must be taken into account in finding the matrix elements
(non-diagonal with respect to configuration included)
of
any physical operator between complex configurations (with any number of
open shells).

\item An idea is proposed and carried out, concerning the most efficient
way to apply the tensor algebra and the quasispin formalism in the most
general case, for the diagonal and the non-diagonal matrix elements as
well, when the bra- and ket- functions have any number of open shells.

\end{enumerate}

The combination of all these improvements allows to efficiently account
for correlation efects practically for any atom or ion of periodical table.

\section*{\bf Acknowledgements}

The author is grateful to Professor Z. Rudzikas for
encouraging and valuable remarks.

\end{document}